\documentclass[preprint,onecolumn,nofootinbib]{revtex4}
\pdfoutput=1
\usepackage[colorlinks=true,linkcolor=blue,urlcolor=blue,filecolor=black,citecolor=red,pdfstartview=FitV,pdftitle={},pdfsubject={},pdfkeywords={},pdfpagemode=None,bookmarksopen=true]{hyperref}
\usepackage{graphicx}
\usepackage{amsmath}
\usepackage{amsfonts}
\usepackage{amssymb}
\usepackage{color}%
\usepackage{dcolumn}
\newcommand{\f}{\begin{equation}}
\newcommand{\ff}{\end{equation}}
\newcommand{\fa}{\begin{eqnarray}}
\newcommand{\ffa}{\end{eqnarray}}

\begin{document}
\title{Momentum dissipation and holographic transport without self-duality}
\author{Jian-Pin Wu $^{1,2}$}
\email{jianpinwu@mail.bnu.edu.cn}
\affiliation{
$^1$ Institute of Gravitation and Cosmology, Department of
Physics, School of Mathematics and Physics, Bohai University, Jinzhou 121013, China\ \\
$^2$ Shanghai Key Laboratory of High Temperature Superconductors,
Shanghai, 200444, China}
\begin{abstract}

We implement the momentum dissipation introduced by spatial linear axionic fields in a holographic model without self-duality,
broke by Weyl tensor coupling to Maxwell field, and study its response.
It is found that for the positive Weyl coupling parameter $\gamma>0$,
the momentum dissipation characterized by parameter $\hat{\alpha}$ drives the boundary conformal field theory (CFT),
in which the conductivity exhibits a peak at low frequency,
into the incoherent metallic phase with a dip, which is away from CFT due to the introduction of axionic fields.
While for $\gamma<0$, an oppositive scenario is found.
Our present model provides a possible route toward the problem that which sign of $\gamma$ is the correct description
of the CFT of boson Hubbard model.
In addition, we also investigate the DC conductivity, diffusion constant and susceptibility.
We find that for each of these observables there is a specific value of $\hat{\alpha}$,
for which these observables are independent of $\gamma$.
Finally, the electromagnetic (EM) duality is also studied
and we find that there is also a specific value of $\hat{\alpha}$,
for which the particle-vortex duality related by the change of the sign of $\gamma$ in the boundary theory
holds better than for other values of $\hat{\alpha}$.

\end{abstract}
\maketitle
\tableofcontents
\section{Introduction}

The transport properties, such as the electrical conductivity, heat conductivity and thermoelectric transport, are great important features of real materials.
For the weakly coupled systems, the frequency dependent conductivity exhibits Drude-like peak at low frequency.
Their collective dynamics is well described by the quantum Boltzmann theory of the quasi-particles with long-lived excitations \cite{Damle:1997rxu}.
While for the strongly coupled systems, the picture of the quasi-particle is absent
and the Boltzmann theory is usually invalid \footnote{When quasi-particle excitations are only weakly
broken, the perturbative method in the Boltzmann framework is developed to deal with such systems, see for example, \cite{Damle:1997rxu,Sachdev:1998,Fritz:2008,WitczakKrempa:2012um}.}.

The anti-de Sitter/conformal field theory (AdS/CFT) correspondence \cite{Maldacena:1997re,Gubser:1998bc,Witten:1998qj,Aharony:1999ti}
provides a powerful tool and novel mechanism to study the transport of the strongly coupled systems.
One of the long-standing important issues in strongly coupling systems is the quantum critical (QC) dynamics described by CFT \cite{Sachdev:QPT} (also refer to \cite{Cha:1991,Damle:1997rxu,Smakov:2005,WitczakKrempa:2012um,Chen:2013ppa,Gazit:2013hga,Gazit:2014,Witczak-Krempa:2015jca}).
A controlled manner in traditional field theory is absent in studying the QC physics at finite temperature.
Also the numerical simulations also suffer from the ``sign'' problem and usually fail.
Here the power of holography is evident.
More recently, remarkable progresses have been made in the study of the transport properties of QC physics by holography \cite{Myers:2010pk,WitczakKrempa:2012gn,WitczakKrempa:2013ht,Witczak-Krempa:2013nua,Witczak-Krempa:2013aea,Katz:2014rla}.
They introduce an extra four-derivative interaction, the Weyl tensor $C_{\mu\nu\rho\sigma}$, coupled to Maxwell field,
in the Schwarzschild-AdS (SS-AdS) black brane, which is dual to a neutral plasma at finite temperature,
to study the transport behavior of the QC physics \footnote{The transport of the Maxwell-Weyl
system in a perturbative higher-derivative neutral background has also been studied in \cite{Bai:2013tfa}.}.
Since the breakdown of the electromagnetic (EM) self-duality, a frequency dependent optical conductivity is observed in this neutral plasma \footnote{Note that due to the EM self-duality,
the optical conductivity is frequency independent in the neutral system dual to the standard Maxwell theory in four dimensional SS-AdS bulk spacetime \cite{Herzog:2007ij}.} \cite{Myers:2010pk}.
In particular, the conductivity at low frequency displays a peak for $\gamma>0$, which resembles the particle excitation described by the Boltzmann theory \cite{Myers:2010pk}.
While for $\gamma<0$, it exhibits a dip and is similar to the vortex case \cite{Myers:2010pk}.
Here $\gamma$ is the parameter controlling the coupling strength of the Weyl term.
It also provides a possible route to access the CFT of the superfluid-insulator
quantum critical point (QCP) described by the boson Hubbard model though there is still a degree
of freedom of the sign of $\gamma$ \footnote{We provide a brief introduction on this problem in Section \ref{sec-optical-conductivity}.
For the more details, please see \cite{Sachdev:2011wg}.} \cite{Sachdev:2011wg}.
Also they find that a particle-vortex duality in the dual boundary field theory,
which is related by the change of the sign of $\gamma$ \cite{Myers:2010pk,WitczakKrempa:2012gn,WitczakKrempa:2013ht,Witczak-Krempa:2013nua,Witczak-Krempa:2013aea,Katz:2014rla}.
Furthermore, based on this framework \cite{Myers:2010pk}, some important results are achieved \cite{WitczakKrempa:2012gn,WitczakKrempa:2013ht,Witczak-Krempa:2013nua,Witczak-Krempa:2013aea,Katz:2014rla}.
For instance, by combining high precision quantum Monte Carlo (QMC) simulations with the results
from the boundary CFT dual to the Maxwell-Weyl system in SS-AdS geometry,
a quantitative description of the transports of QC physics without quasi-particle excitation is built \cite{Witczak-Krempa:2013nua,Katz:2014rla},
which is experimentally testable.

In this paper, we intend to implement the momentum dissipation into the Maxwell-Weyl system studied in \cite{Myers:2010pk,WitczakKrempa:2012gn,WitczakKrempa:2013ht,Witczak-Krempa:2013nua,Witczak-Krempa:2013aea,Katz:2014rla}
and investigate its response. There are many ways to implement the momentum dissipation in holographic manner,
see for example \cite{Horowitz:2012ky,Horowitz:2012gs,Ling:2013nxa,Donos:2012js,Donos:2013eha,Donos:2014uba,Vegh:2013sk,Andrade:2013gsa},
and many interesting results have been obtained, for example \cite{Blake:2013bqa,Blake:2013owa,Ling:2016wyr,Ling:2015dma,Ling:2014laa,Ling:2014saa,Zeng:2014uoa,Mozaffara:2016iwm}.
Here we adopt a simple way proposed in \cite{Andrade:2013gsa},
where the momentum dissipation is implemented by a pair of massless field, $\Phi_I$ with $I=1,2$, which are
spatial linear dependent in bulk.
It is also referred as ``mean-field disordered'' \cite{Grozdanov:2015qia} due to the homogeneous background geometry.
Note that $\Phi_I$ correspond to turning on spatial linear sources in the dual boundary theory, \emph{i.e.},
\fa
\phi_I^{(0)}\propto \alpha x_I\,,
\ffa
with $\alpha$ being constant.
This nonuniform source means that a dimensionful parameter, \emph{i.e.}, $\alpha$, is introduced into the dual boundary theory
and so the physics we studying is that away from QCP.
We hope that our present model provides wider route to address
whether the excitation of the CFT of the superfluid-insulator QCP described by
the boson Hubbard model is particle-like or vortex-like
and also toward the problem that which sign of $\gamma$ is the correct description of this CFT.
In addition, the proximity effect in QCP, which also alters some observables such as the optical conductivity,
is also important and has been explored in \cite{Sachdev:QPT,Cha:1991,Damle:1997rxu,Smakov:2005,WitczakKrempa:2012um,Chen:2013ppa,Gazit:2013hga,Gazit:2014,Witczak-Krempa:2015jca,Lucas:2016fju}.
Our present work will also provide some insight into the transport properties away from QCP in holographic framework.

Our paper organizes as follows.
We begin with a review of the holographic framework without EM self-duality in Section \ref{sec-HF}.
We then introduce a neutral axionic theory, which is responsible for the momentum dissipation in Section \ref{neuaxion}.
The optical conductivity of the boundary field theory dual to the Maxwell-Wely system in the neutral axionic geometry is studied in \ref{sec-optical-conductivity}.
We mainly focus on the role the momentum dissipation plays in the transport propertie in our present model.
We also study the diffusion constant and susceptibility of the dual boundary field theory in Section \ref{sec-discussion}.
In Section \ref{sec-EM-duality}, we discuss the EM duality.
We conclude with a brief discussion and some open questions in Section \ref{sec-discussion}.
In Appendix \ref{Bounds}, we discuss the constraints imposing on the Weyl coupling parameter $\gamma$ in the neutral axionic geometry due to the causality and the instabilities.

\section{Holographic framework without EM self-duality}\label{sec-HF}

The optical conductivity in the neutral plasma dual to the standard Maxwell theory in four dimensional AdS spacetimes
is frequency independent due to the EM self-duality \cite{Herzog:2007ij}.
To have a frequency dependent optical conductivity in the neutral plasma, we need to break the EM self-duality.
A simple way is to introduce the Weyl tensor $C_{\mu\nu\rho\sigma}$ coupled to gauge field as
\cite{Myers:2010pk,WitczakKrempa:2012gn,WitczakKrempa:2013ht,Witczak-Krempa:2013nua,Witczak-Krempa:2013aea,Katz:2014rla,Bai:2013tfa,Ritz:2008kh}
\fa
\label{ac-ma}
S_1=\frac{1}{g_F^2}\int d^4x\sqrt{-g}\Big(-\frac{1}{4}F_{\mu\nu}F^{\mu\nu}+\gamma C_{\mu\nu\rho\sigma}F^{\mu\nu}F^{\rho\sigma}\Big)\,,
\ffa
where $F=dA$ is the curvature of gauge field $A$ and $g_F^2$ is an effective dimensionless gauge coupling,
which shall be set $g_F=1$ in the numerical calculation.
In this theory, there is a crucial dimensionless coupling parameter $\gamma$,
which controls the coupling strength of the Maxwell-Weyl term.
The Weyl term is a specific combination of some four-derivative interaction term \cite{Myers:2010pk},
which can be expected to emerge as quantum corrections in the low energy effective action in string theory context \cite{Hanaki:2006pj,Cremonini:2008tw}.

It is more convenient for subsequent calculations and discussions to
write down the action (\ref{ac-ma}) in a general form \cite{Myers:2010pk}
(also see \cite{WitczakKrempa:2012gn,WitczakKrempa:2013ht,Witczak-Krempa:2013nua,Witczak-Krempa:2013aea,Katz:2014rla})
\fa
\label{ac-SA}
S_A=\int d^4x\sqrt{-g}\Big(-\frac{1}{8g_F^2}F_{\mu\nu}X^{\mu\nu\rho\sigma}F_{\rho\sigma}\Big)\,.
\ffa
A new tensor $X$ is introduced in the above equation as
\fa
X_{\mu\nu}^{\ \ \rho\sigma}=I_{\mu\nu}^{\ \ \rho\sigma}-8\gamma C_{\mu\nu}^{\ \ \rho\sigma}\,,
\label{X-tensor}
\ffa
with an identity matrix acting on two-forms
\fa
I_{\mu\nu}^{\ \ \rho\sigma}=\delta_{\mu}^{\ \rho}\delta_{\nu}^{\ \sigma}-\delta_{\mu}^{\ \sigma}\delta_{\nu}^{\ \rho}\,.
\label{I-tensor}
\ffa
It is easy to find that the $X$ tensor possess the following symmetries
\fa
X_{\mu\nu\rho\sigma}=X_{[\mu\nu][\rho\sigma]}=X_{\rho\sigma\mu\nu}\,.
\label{X-sym}
\ffa
When we set $X_{\mu\nu}^{\ \ \rho\sigma}=I_{\mu\nu}^{\ \ \rho\sigma}$, the theory (\ref{ac-ma}) reduces to the standard Maxwell theory.
And then, from the action (\ref{ac-SA}), we have the equation of motion as
\fa
\nabla_{\nu}(X^{\mu\nu\rho\sigma}F_{\rho\sigma})=0\,.
\label{eom-Max}
\ffa

In presence of the Weyl term, the EM self-duality breaks down \cite{Myers:2010pk}.
However, we can still construct the dual EM theory for the gauge theory (\ref{ac-ma}).
For more details, we can see \cite{Myers:2010pk}. Here we directly write down the corresponding dual EM theory
\fa
\label{ac-SB}
S_B=\int d^4x\sqrt{-g}\Big(-\frac{1}{8\hat{g}_F}G_{\mu\nu}\widehat{X}^{\mu\nu\rho\sigma}G_{\rho\sigma}\Big)\,,
\ffa
where $\hat{g}_F^2\equiv 1/g_F^2$ and $G_{\mu\nu}\equiv\partial_{\mu}B_{\nu}-\partial_{\nu}B_{\mu}$.
In addition, the tensor $\widehat{X}$ is defined by
\fa
\widehat{X}_{\mu\nu}^{\ \ \rho\sigma}=-\frac{1}{4}\varepsilon_{\mu\nu}^{\ \ \alpha\beta}(X^{-1})_{\alpha\beta}^{\ \ \gamma\lambda}\varepsilon_{\gamma\lambda}^{\ \ \rho\sigma}\,,
\label{X-hat}
\ffa
where $\varepsilon_{\mu\nu\rho\sigma}$ is volume element and $X^{-1}$ is defined by
\fa
\frac{1}{2}(X^{-1})_{\mu\nu}^{\ \ \rho\sigma}X_{\rho\sigma}^{\ \ \alpha\beta}\equiv I_{\mu\nu}^{\ \ \alpha\beta}\,.
\label{X-ne-def}
\ffa
Also we can derive the equation of motion of the dual theory (\ref{ac-SB}) as
\fa
\nabla_{\nu}(\widehat{X}^{\mu\nu\rho\sigma}G_{\rho\sigma})=0\,.
\label{eom-Max-B}
\ffa

For the standard four-dimensional Maxwell theory, $\widehat{X}_{\mu\nu}^{\ \ \rho\sigma}=I_{\mu\nu}^{\ \ \rho\sigma}$ and therefore the theory (\ref{ac-SA}) and (\ref{ac-SB})
are identical, which means that the Maxwell theory is self-dual.
When the Weyl term is introduced and for small $\gamma$, we find that
\fa
&&
(X^{-1})_{\mu\nu}^{\ \ \rho\sigma}=I_{\mu\nu}^{\ \ \rho\sigma}+8\gamma C_{\mu\nu}^{\ \ \rho\sigma}+\mathcal{O}(\gamma^2)\,,
\label{Xin}
\\
&&
\widehat{X}_{\mu\nu}^{\ \ \rho\sigma}=(X^{-1})_{\mu\nu}^{\ \ \rho\sigma}+\mathcal{O}(\gamma^2)\,£¬
\ffa
which implies that the self-dual is violated for the theory (\ref{ac-ma}) but for small $\gamma$, there is a duality between the actions (\ref{ac-SA}) and (\ref{ac-SB})
with the change of the sign of $\gamma$.

\section{A neutral axionic theory}\label{neuaxion}

We intend to implement the momentum dissipation in the Maxwell-Weyl system and study its response.
The simplest way is to introduce a pair of spatial linear dependent axionic fields \cite{Andrade:2013gsa},
in which the action is
\fa
\label{ac-ax}
S_0=\int d^4x\sqrt{-g}\Big(R+6-\frac{1}{2}\sum_{I=x,y}(\partial \phi_I)^2\Big)
\,,
\ffa
where $\phi_I=\alpha x_I$ with $I=x,y$ and $\alpha$ being a constant.
In this action, there is a negative cosmological constant $\Lambda=-6$, which supports an asymptotically AdS spacetimes \footnote{Here,
without loss of generality we have set the AdS radius $L=1$ for simplify.}.
When the momentum dissipation is weak,
the standard Maxwell theory with action (\ref{ac-ax}) describes coherent metallic behavior \cite{Andrade:2013gsa,Kim:2014bza,Davison:2014lua}.
Conversely once the momentum dissipation is strong, we have an incoherent metal \cite{Andrade:2013gsa,Kim:2014bza,Davison:2014lua}.

Since the Einstein-Maxwell-axion-Weyl (EMA-Weyl) theory (Eqs.(\ref{ac-ax}) and (\ref{ac-ma})) involves solving a set of
third order nonlinear differential equations, it is hard to solve them even numerically.
As an alternative method, we can construct analytical background solutions up to the first order of
the Weyl coupling parameter $\gamma$ \cite{Myers:2009ij,Liu:2008kt,Cai:2011uh,Dey:2015poa,Dey:2015ytd,Mahapatra:2016dae,Ling:2016dck}.
But it is still hard to obtain the frequency dependent conductivity and only the DC conductivity is worked out in \cite{Ling:2016dck}.
As the first step, here we shall follow the strategy in \cite{Myers:2010pk,WitczakKrempa:2012gn,WitczakKrempa:2013ht,Witczak-Krempa:2013nua,Witczak-Krempa:2013aea,Katz:2014rla,Bai:2013tfa}
and turn to study the transports of the Maxwell-Weyl system (\ref{ac-ma})
in the neutral plasma dual to the Einstein-axions (EA) theory (\ref{ac-ax}).

The neutral black brane solution of the EA action (\ref{ac-ax}) can be written down as \cite{Andrade:2013gsa}
\fa
\label{bl-br}
ds^2=\frac{1}{u^2}\Big(-f(u)dt^2+\frac{1}{f(u)}du^2+dx^2+dy^2\Big)\,,
\ffa
where
\fa
\label{fu}
f(u)=(1-u)p(u)\,,~~~~~~~
p(u)=\Big(1-\frac{\alpha^2}{2}\Big)u^2+u+1\,.
\ffa
$u=0$ is the asymptotically AdS boundary while the horizon locates at $u=1$.
And then, the Hawking temperature can be expressed as
\fa
\label{ha-te}
T=\frac{p(1)}{4\pi}=\frac{3-\frac{\alpha^2}{2}}{4\pi}\,.
\ffa
Since the black brane solution (\ref{bl-br}) with (\ref{fu}) is only parameterized by one scaling-invariant parameter $\hat{\alpha}=\alpha/4\pi T$,
for later convenience, we reexpress the function $p(u)$ as
\fa
\label{pu}
p(u)=\frac{\sqrt{1+6\hat{\alpha}^2}-2\hat{\alpha}^2-1}{\hat{\alpha}^2}u^2+u+1\,.
\ffa
Further, the energy density $\epsilon$, pressure $p$ and entropy density $s$ of the dual boundary theory can be calculated as \cite{Andrade:2013gsa}
\fa
\epsilon=2\Big(1-\frac{\alpha^2}{2}\Big)\,,\,\,\,\, p=1+\frac{\alpha^2}{2}\,,\,\,\,\, s=4\pi\,.
\ffa

At this moment, there are some comments presenting in order.
First, it is easy to see that from (\ref{ha-te}) at zero temperature ($\alpha=\sqrt{6}$),
the IR geometry is AdS$_2\times \mathbb{R}_2$, which is similar to the Reissner-Nordstr$\ddot{o}$m-AdS (RN-AdS) black brane.
Second, there is special value of $\alpha=\sqrt{2}$ where the energy density $\epsilon$ vanishes.
At this point, there is self-duality in Maxwell equations and the AC heat conductivity is frequency-independent \cite{Davison:2014lua}.
Third, the axionic fields $\phi_I$ in bulk correspond to
turning on sources in the dual boundary theory which is linearly dependent of the spatial coordinate.
Such sources result in the momentum dissipation, which is controlled by the parameter $\hat{\alpha}$.
At the same time, as pointed out in the introduction,
$\Phi_I$ introduces the nonuniform source with a dimensionful parameter $\alpha$ in the dual boundary theory
such that the system we are studying is away from QCP.

\section{Optical conductivity}\label{sec-optical-conductivity}

A simple but important transport behavior is the electrical optical conductivity at zero momentum.
We mainly study it in this paper.
The other transport properties, such as the thermal conductivity, the optical conductivity at finite momentum,
will be studied elsewhere.

In holographic framework, the optical conductivity along $y$-direction can be calculated by \footnote{Due to the symmetry between $x$ and $y$ directions,
we only need calculate the conductivity along either $x$ or $y$ direction.}
\fa
\sigma(\omega)=\frac{\partial_uA_y(u,\hat{\omega},\hat{q}=0)}{i\omega A_y(u,\hat{\omega},\hat{q}=0)}\,.
\label{con-def}
\ffa
$A_y(u,\hat{\omega},\hat{q}=0)$ is the perturbation of the gauge field at zero momentum along $y$-direction in Fourier space (see Eq.(\ref{A-Fourier}) in Appendix \ref{Bounds}).
For $\hat{q}=0$, all the perturbative equations of the gauge field (Eqs.(\ref{Ma-AtI})-(\ref{Ma-Ay}) in Appendix \ref{Bounds}) decouple.
Therefore, we only need to solve Eq.(\ref{Ma-Ay}) to obtain the optical conductivity.
Note that $\hat{\omega}$ and momentum $\hat{q}$ are the dimensionless frequency and momentum, respectively,
which have been defined in Eq.(\ref{hat-omega-q}) in Appendix \ref{Bounds}.

\subsection{Optical conductivity}

In \cite{Myers:2010pk,WitczakKrempa:2012gn,WitczakKrempa:2013ht,Witczak-Krempa:2013nua,Witczak-Krempa:2013aea,Katz:2014rla},
the optical conductivity of the boundary field theory dual to SS-AdS geometry has been explored.
For $\gamma>0$, it exhibits a peak at low frequency\footnote{There is a deviation from the standard Drude formula for $\gamma\in\mathcal{S}_0$.
We shall illustrate this point below.},
which qualitatively resembles the Boltzmann transport of particles.
While for $\gamma<0$, a dip appears, which is similar to the excitation of vortices.
It provides a possible route to resolve whether the excitation of the CFT of the superfluid-insulator QCP described by
the boson Hubbard model is particle-like or vortex-like.
Before proceeding, let us briefly address this problem. For the details, we can refer to \cite{Sachdev:2011wg}.
In the insulating phase of the boson Hubbard model, it is the excitation of the particle and hole and
so we can infer that the conductivity at low frequency should exhibit a peak if we approach the QCP from the insulator side.
However, if we approach the QCP form the superfluid side, which described by the excitation of the vortices, the conductivity at low frequency should be a dip.
Until now, we do not know which of the two qualitatively distinct results is correct.
The Maxwell-Weyl system in the SS-AdS geometry provides a possible description for the CFT of the boson Hubbard model in holographic framework
though we still have a degree of freedom of the sign of $\gamma$.
Here, we hope that implementing the momentum dissipation, which is also equivalent to
the ``mean-field disordered'' effect \cite{Grozdanov:2015qia},
will provide more clues in addressing this problem.

\begin{figure}
\center{
\includegraphics[scale=0.6]{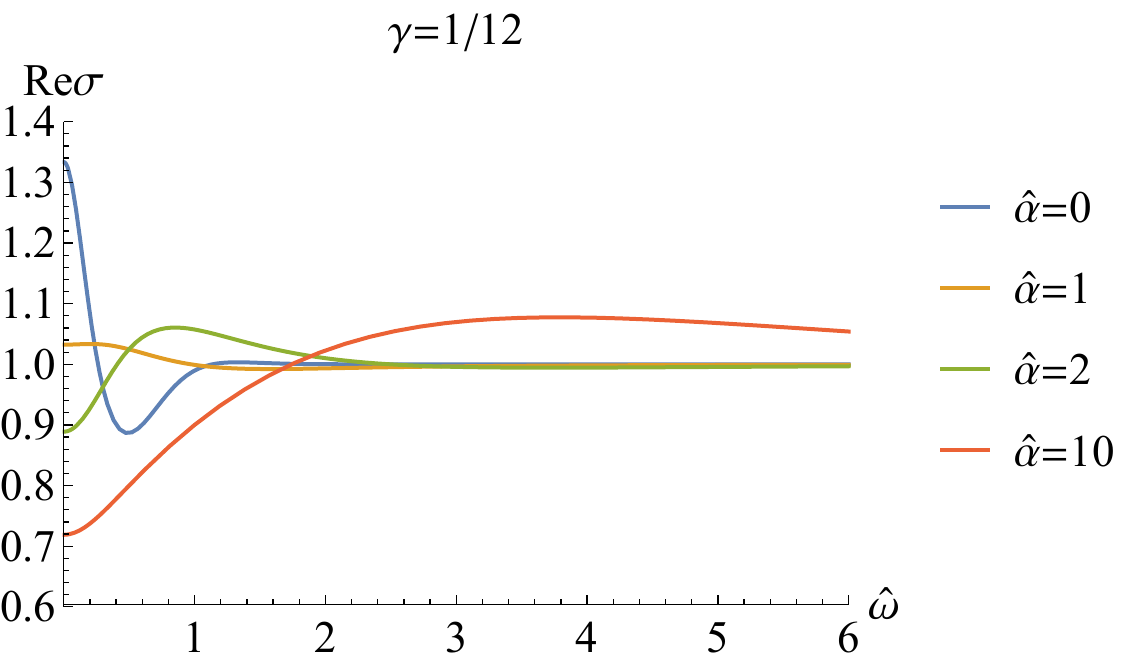}\ \hspace{0.8cm}
\includegraphics[scale=0.6]{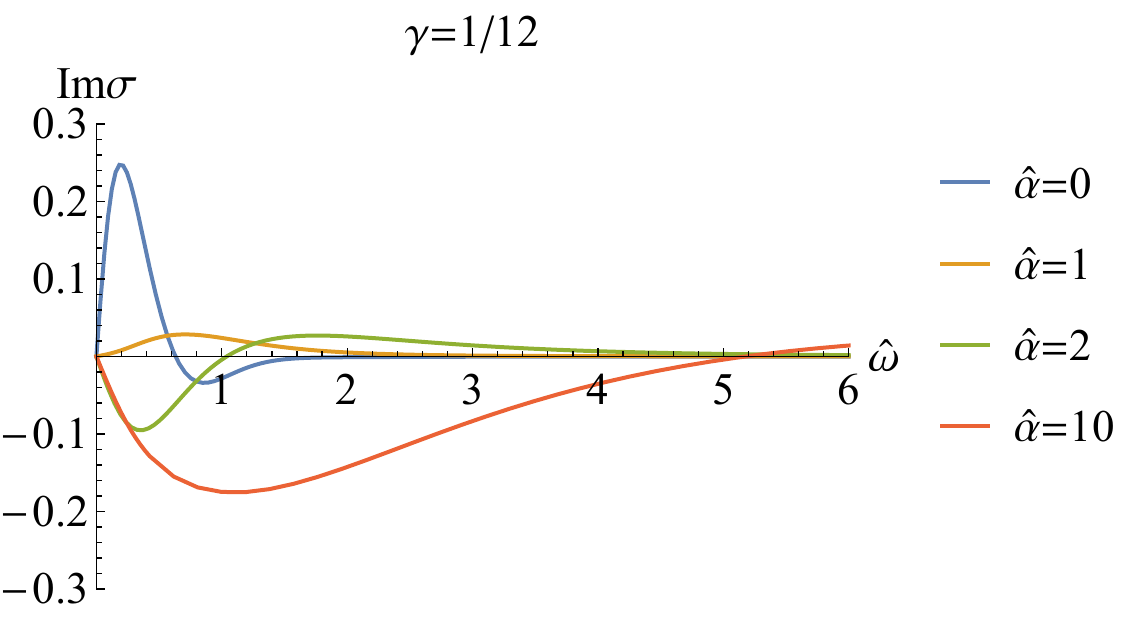}\ \\
\includegraphics[scale=0.6]{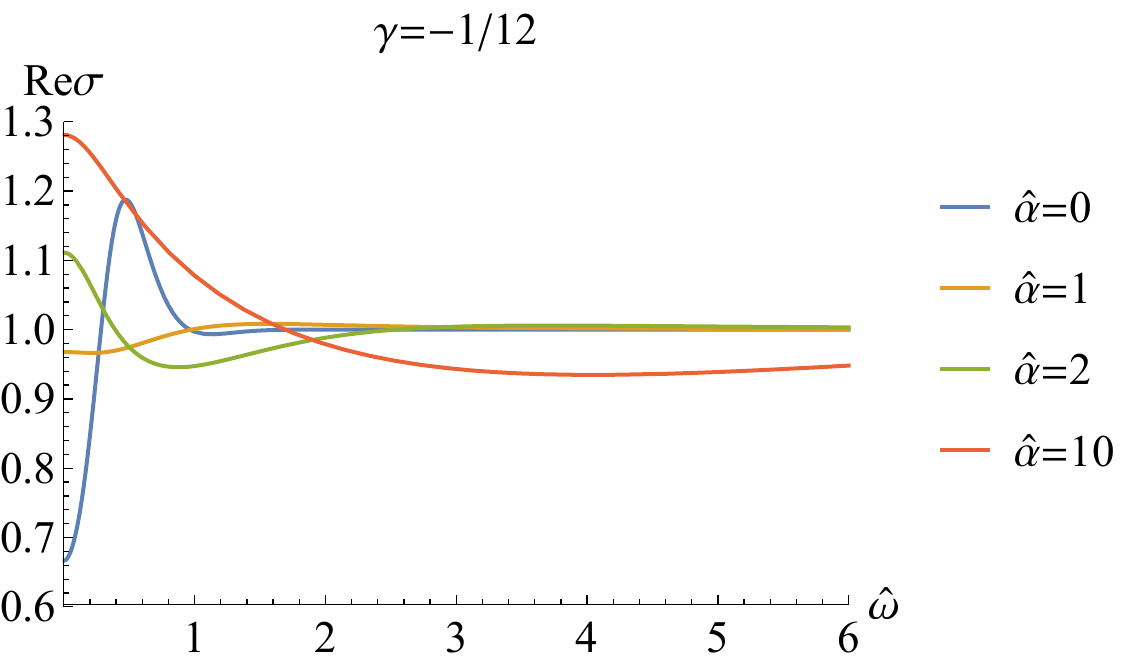}\ \hspace{0.8cm}
\includegraphics[scale=0.6]{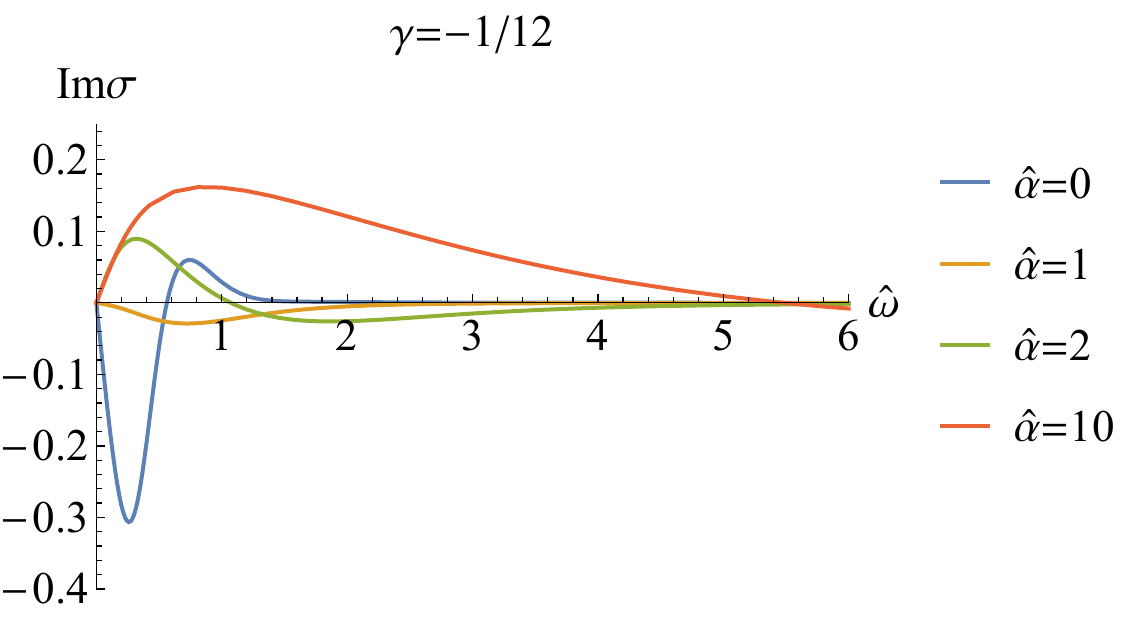}\ \\
\caption{\label{fig-con} The optical conductivity $\sigma(\hat{\omega})$ as the function of $\hat{\omega}$
with different $\hat{\alpha}$ for fixed $\beta$ (the plots above is for $\gamma=1/12$ and the one below for $\gamma=-1/12$).}}
\end{figure}

Now let us see what happens when the momentum dissipation is implemented in the Maxwell-Weyl system.
The top plots of FIG.\ref{fig-con} show the real and imaginary part of the optical conductivity $\sigma(\hat{\omega})$
for $\gamma=1/12$ and different $\hat{\alpha}$.
We observe that at small $\hat{\alpha}$, a peak displays in the low frequency optical conductivity,
and gradually degrades as $\hat{\alpha}$ increases, eventually becomes a dip.
It means that the disorder effect induced by axions drives the CFT described by Maxwell-Weyl system with positive $\gamma$
into the incoherent metallic phase with a dip.
While for $\gamma<0$, an oppositive scenario is found (see the bottom plots of FIG.\ref{fig-con}).
That is to say, as $\hat{\alpha}$ increases, the dip in optical conductivity at low frequency
gradually upgrades and eventually develops into a peak.
It indicates that if the CFT is described by Maxwell-Weyl system with negative $\gamma$,
then the disorder drives it into the metallic phase characterized by a peak.

Though the present model still cannot give a definite answer for which sign of $\gamma$ being the correct description of the CFT of boson Hubbard model,
it indeed provide a route toward this problem, which can be detected in future condensed matter and ultracold atomic gases experiments
or solved in theory by introducing the disorder effect into the boson Hubbard model.
In holographic framework, we can introduce different types of disorder, for example the Q-lattice \cite{Donos:2013eha,Donos:2014uba},
in the Maxwell-Weyl system to see how universal results presented here are.
We shall address this problem in near future.
Next, we present more details on the optical conductivity of our present model, in particular its low frequency behavior.

\subsection{The low frequency behavior of the optical conductivity}

In this subsection, we intend to study the low frequency behavior of the optical conductivity
and try to give some insights into the coherent/incoherent behavior from the momentum dissipation and the Weyl term.

\begin{figure}
\center{
\includegraphics[scale=0.55]{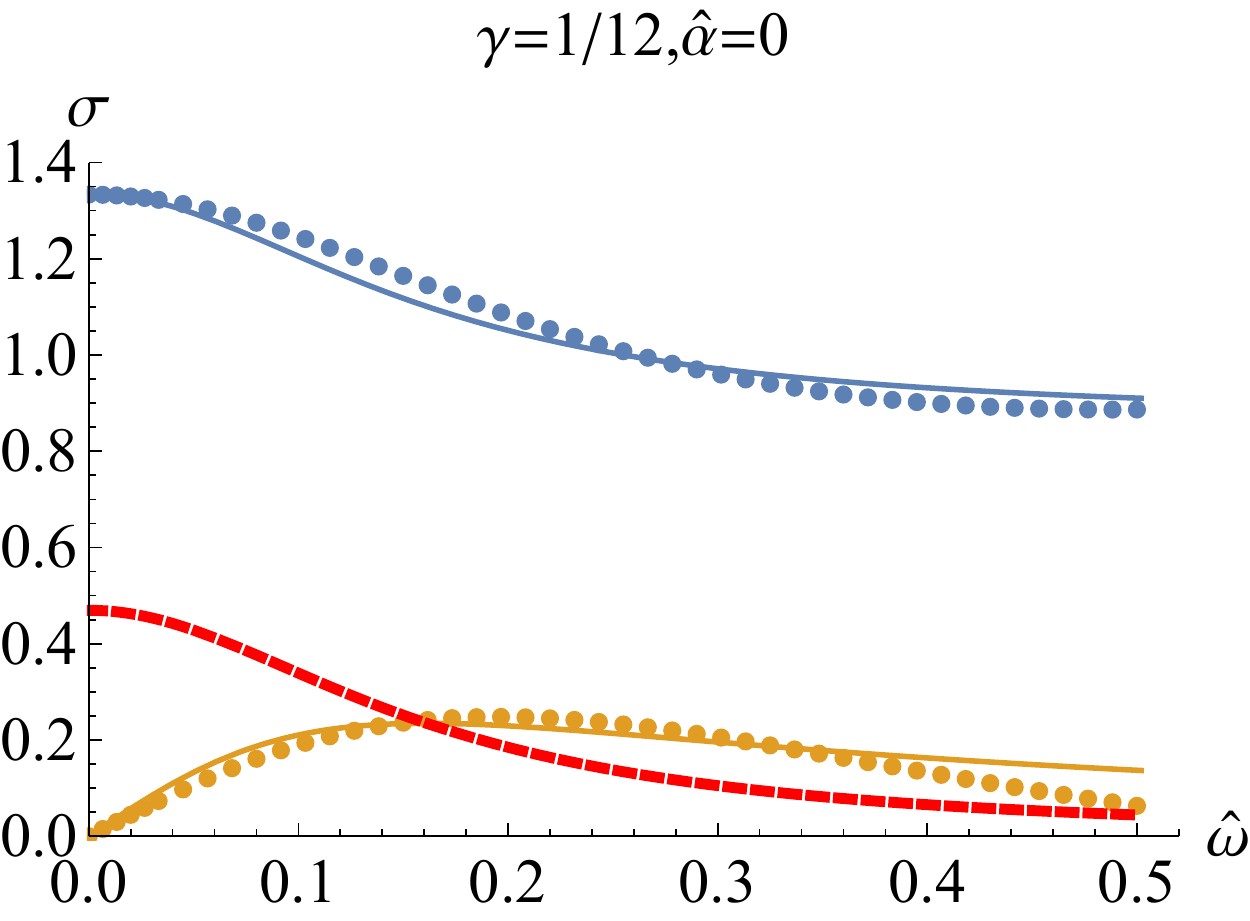}\ \hspace{1cm}
\includegraphics[scale=0.55]{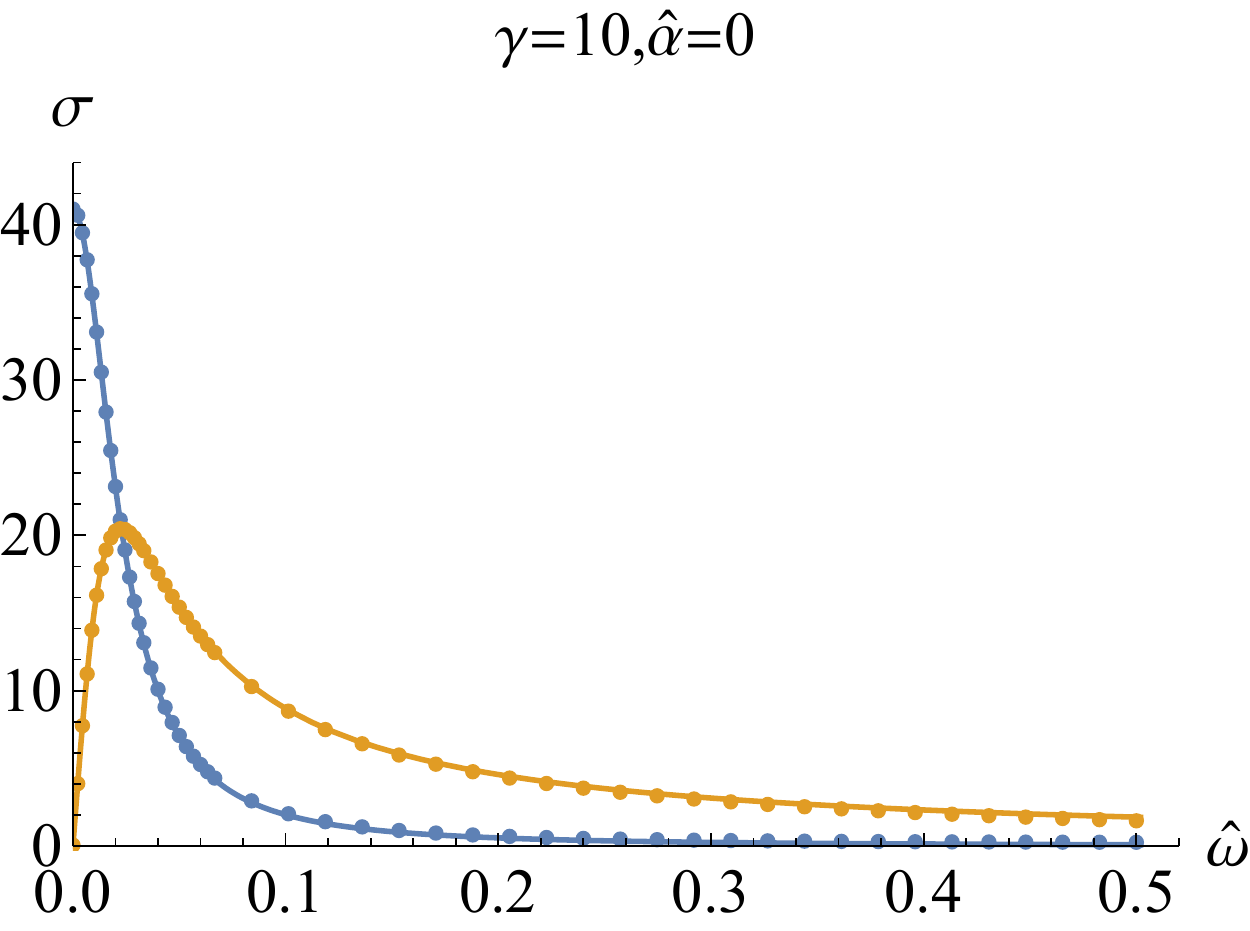}\ \\
\caption{\label{fig-drude} The low frequency behavior of the optical conductivity without momentum dissipation for $\gamma=1/12$ (left plot) and $\gamma=10$ (right plot).
The left plot is fitted by a modified Drude formula (\ref{modified-Drude}). The red dashed line in left plot is the real part of optical conductivity when fitted with the standard Drude formula.
The right plot is fitted by the standard Drude formula.}}
\end{figure}

In the boundary field theory dual to the Maxwell-Weyl system in SS-AdS geometry,
although a peak emerges in the optical conductivity at low frequency for $\gamma>0$ \cite{Myers:2010pk},
there is in fact a deviation from the standard Drude formula if we require $\gamma\in\mathcal{S}_0$.
Inspired by the incoherent metallic phase studied in \cite{Hartnoll:2014lpa}
(also see \cite{Kim:2014bza,Ge:2014aza,Davison:2015taa,Davison:2015bea,Ling:2015exa,Zhou:2015qui}
for the related studies),
we use the following modified Drude formula to fit the data for $\gamma=1/12$,
\fa
\sigma(\hat{\omega})=\frac{K\tau}{1-i\hat{\omega}\tau}+\sigma_Q\,,
\label{modified-Drude}
\ffa
where $K$ is a constant, $\tau$ the relaxation time and $\sigma_Q$ characters the incoherent degree.
The left plot in FIG.\ref{fig-drude} exhibits such incoherent non-Drude behavior with $\sigma_Q=0.866$.
If we relax $\gamma$ such that it is well beyond the upper bound, \emph{i.e.}, $\gamma\gg1$,
we shall have a coherent Drude behavior in the low frequency optical conductivity (right plot in FIG.\ref{fig-drude}).
But it violates the causality and there may be some instabilities and thus we do not discuss such case.

\begin{figure}
\center{
\includegraphics[scale=0.55]{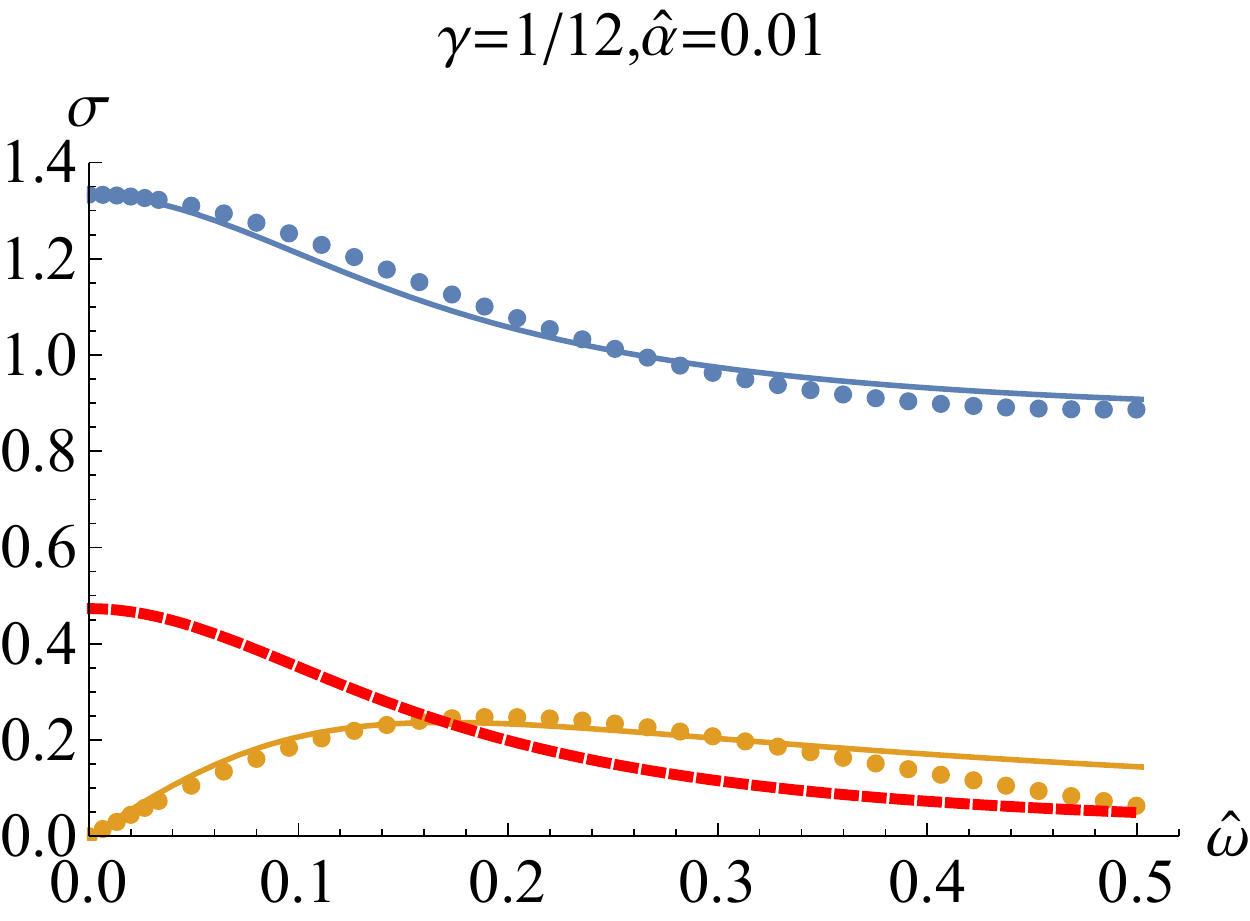}\ \hspace{1cm}
\includegraphics[scale=0.55]{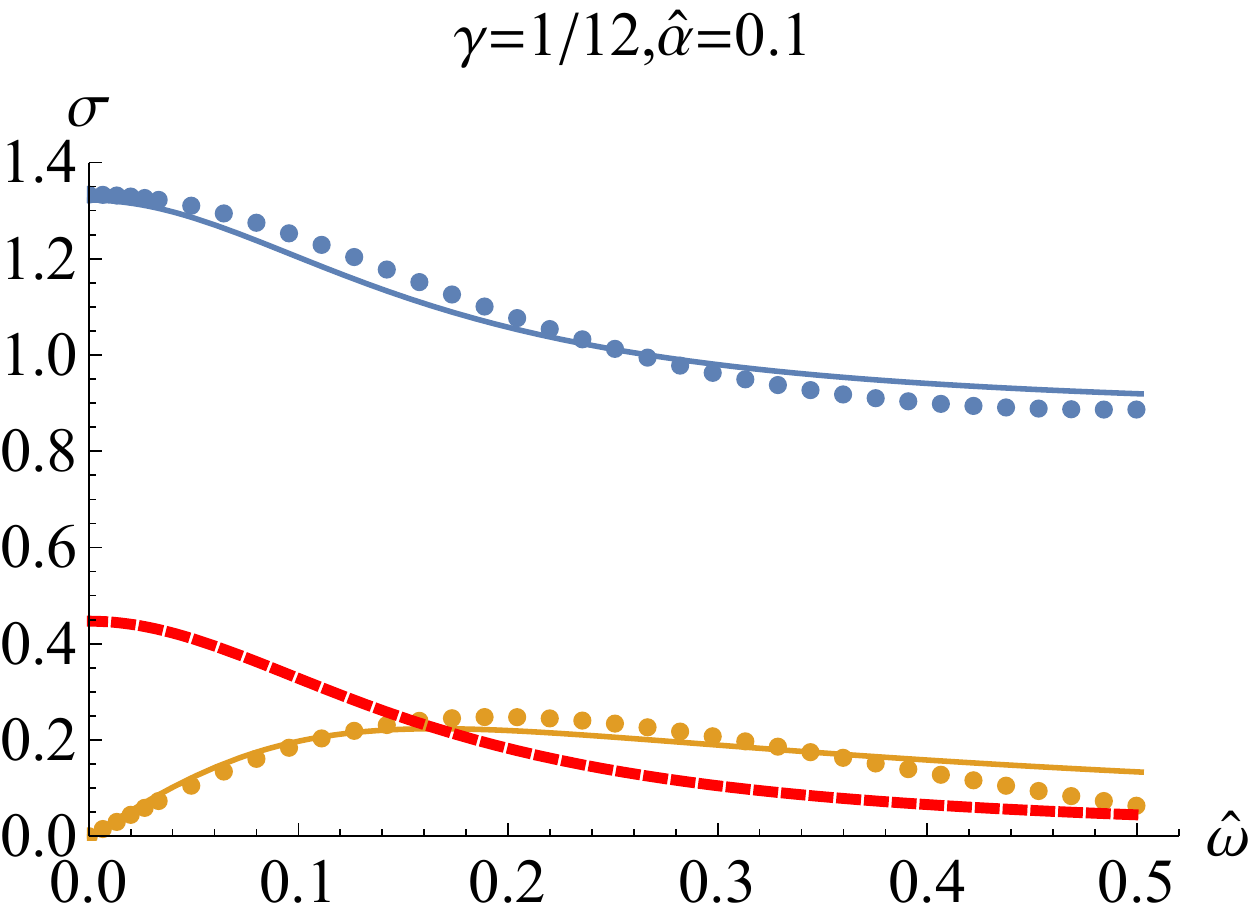}\ \\
\includegraphics[scale=0.55]{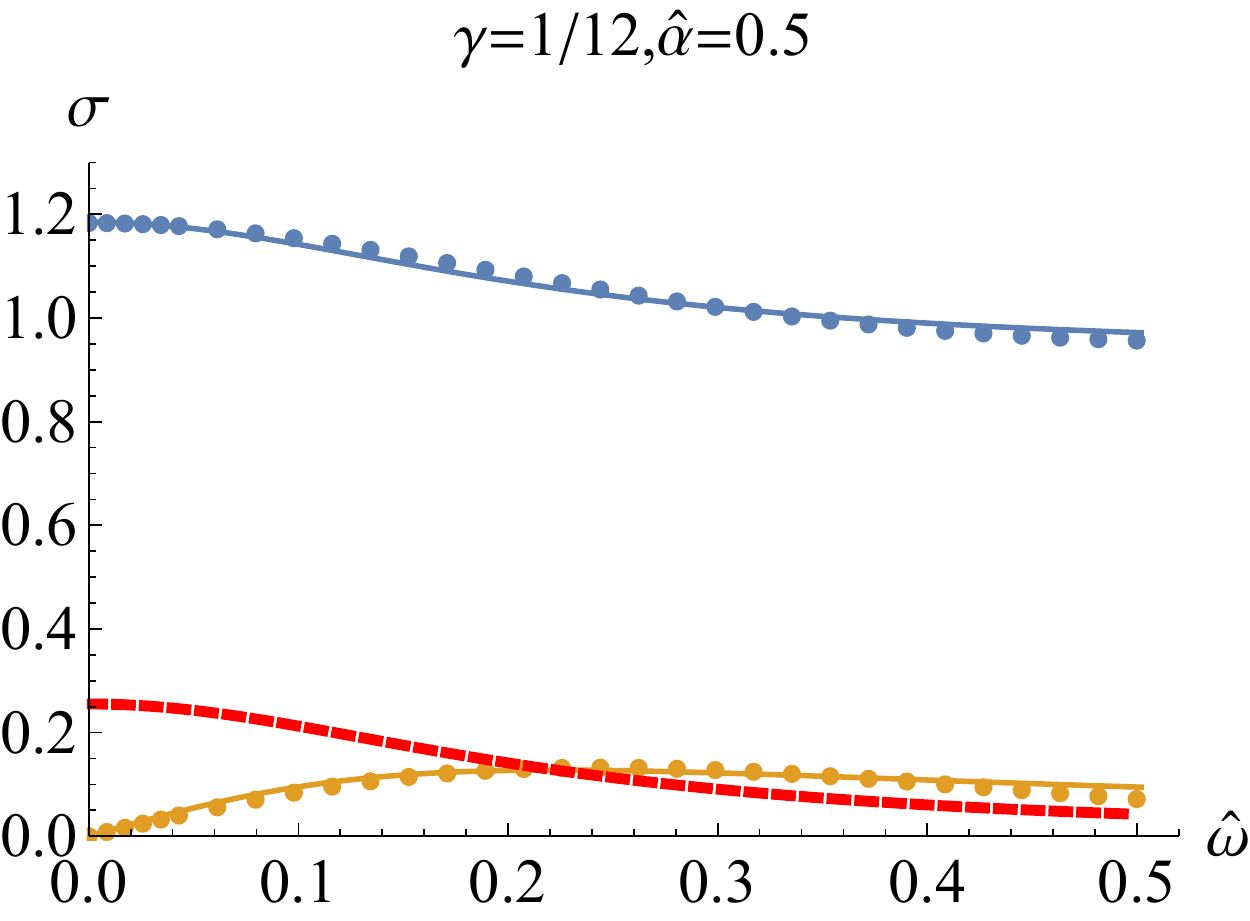}\ \hspace{1cm}
\includegraphics[scale=0.55]{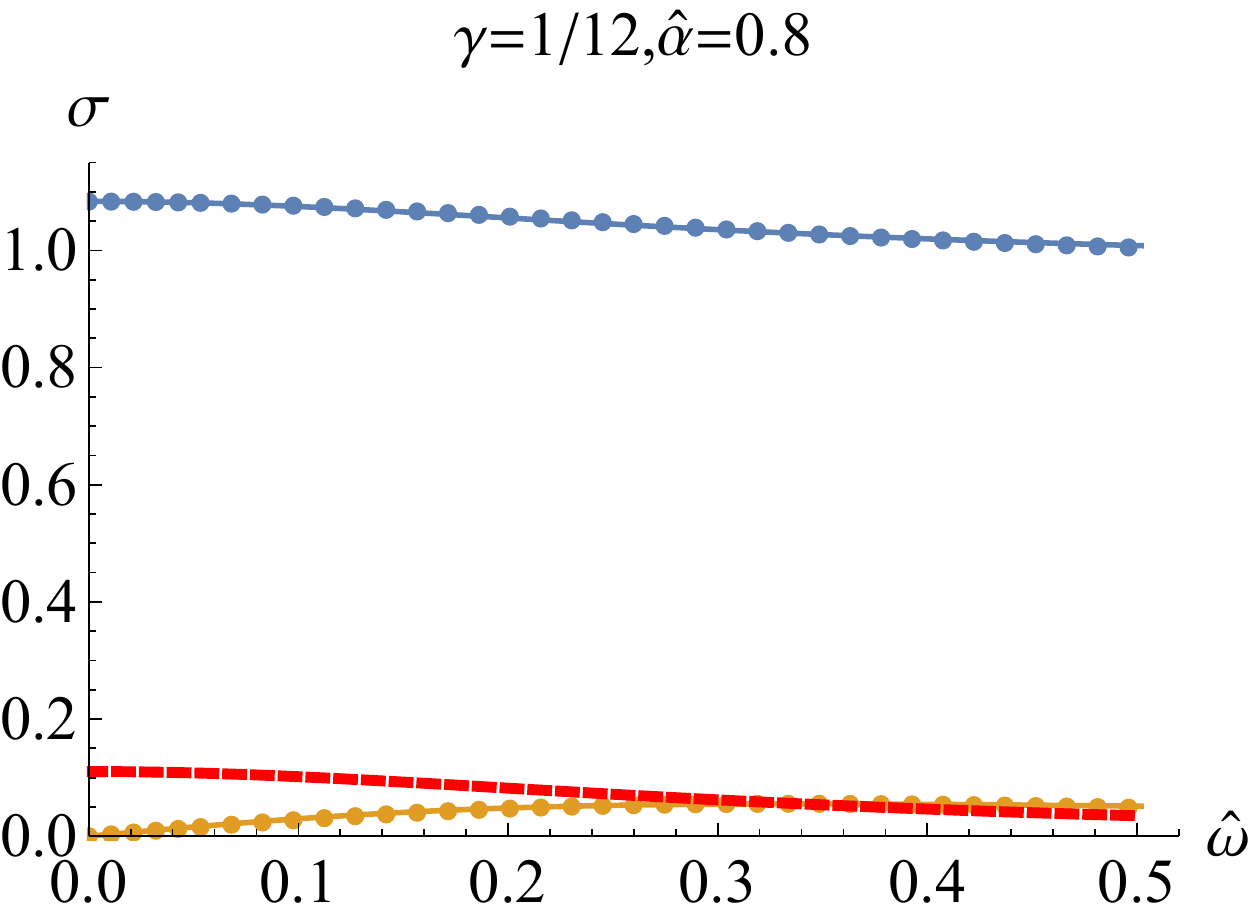}\ \\
\caption{\label{fig-drude_p_gamma_a} The low frequency behavior of the optical conductivity with momentum dissipation for different $\hat{\alpha}$ and $\gamma=1/12$.
They are fitted by a modified Drude formula (\ref{modified-Drude}).
The red dashed line is the real part of optical conductivity fitted with the standard Drude formula.
}}
\end{figure}

Quantitatively, using the modified Drude formula (\ref{modified-Drude}), we fit the low frequency behavior of the optical conductivity
for $\gamma=1/12$ and different $\hat{\alpha}$ (FIG.\ref{fig-drude_p_gamma_a}).
Also, we list the characteristic quantity of incoherence $\sigma_Q$ in Table \ref{table-mod-Drude-p}.
Two illuminating results are summarized as what follows.
First, $\sigma_Q$ increases with the increase of $\hat{\alpha}$ in our present model (Table \ref{table-mod-Drude-p}),
which indicates that the incoherent behavior becomes more evident.
But we also note that for small momentum dissipation ($\hat{\alpha}<0.06$),
$\sigma_Q$ is smaller than that in SS-AdS.
It is because the small momentum dissipation produces coherent contribution,
which reduces the incoherent part from the Weyl term.
While with the increase of $\hat{\alpha}$,
the momentum dissipation becomes strong, combining with that from the Weyl term,
and so the system exhibits more prominent incoherent behavior.
Second, the fit by Eq.(\ref{modified-Drude}) for large $\hat{\alpha}$ is better than that for small $\hat{\alpha}$ (FIG.\ref{fig-drude_p_gamma_a}).
It is because for small $\hat{\alpha}$, the incoherent contribution mainly comes from the Weyl term and also implies that
we need a new non-Drude formula beyond the simple case (\ref{modified-Drude}) to depict meticulously the incoherent contribution from the Weyl term.
We shall further explore this question in future.
\begin{widetext}
\begin{table}[ht]
\begin{center}
\begin{tabular}{|c|c|c|c|c|c|c|c|c|}
         \hline
~$\hat{\alpha}$~ &~$0$~&~$0.01$~&~$0.05$~&~$0.1$~&~$0.5$~&~$0.8$~
          \\
        \hline
~$\sigma_Q$~ & ~$0.866$~&~$0.859$~&~$0.865$~&~$0.875$~ & ~$0.929$~&~$0.973$~
          \\
        \hline
\end{tabular}
\caption{\label{table-mod-Drude-p}The characteristic quantity of incoherence $\sigma_Q$ for $\gamma=1/12$ and different $\hat{\alpha}$.}
\end{center}
\end{table}
\end{widetext}

\begin{figure}
\center{
\includegraphics[scale=0.6]{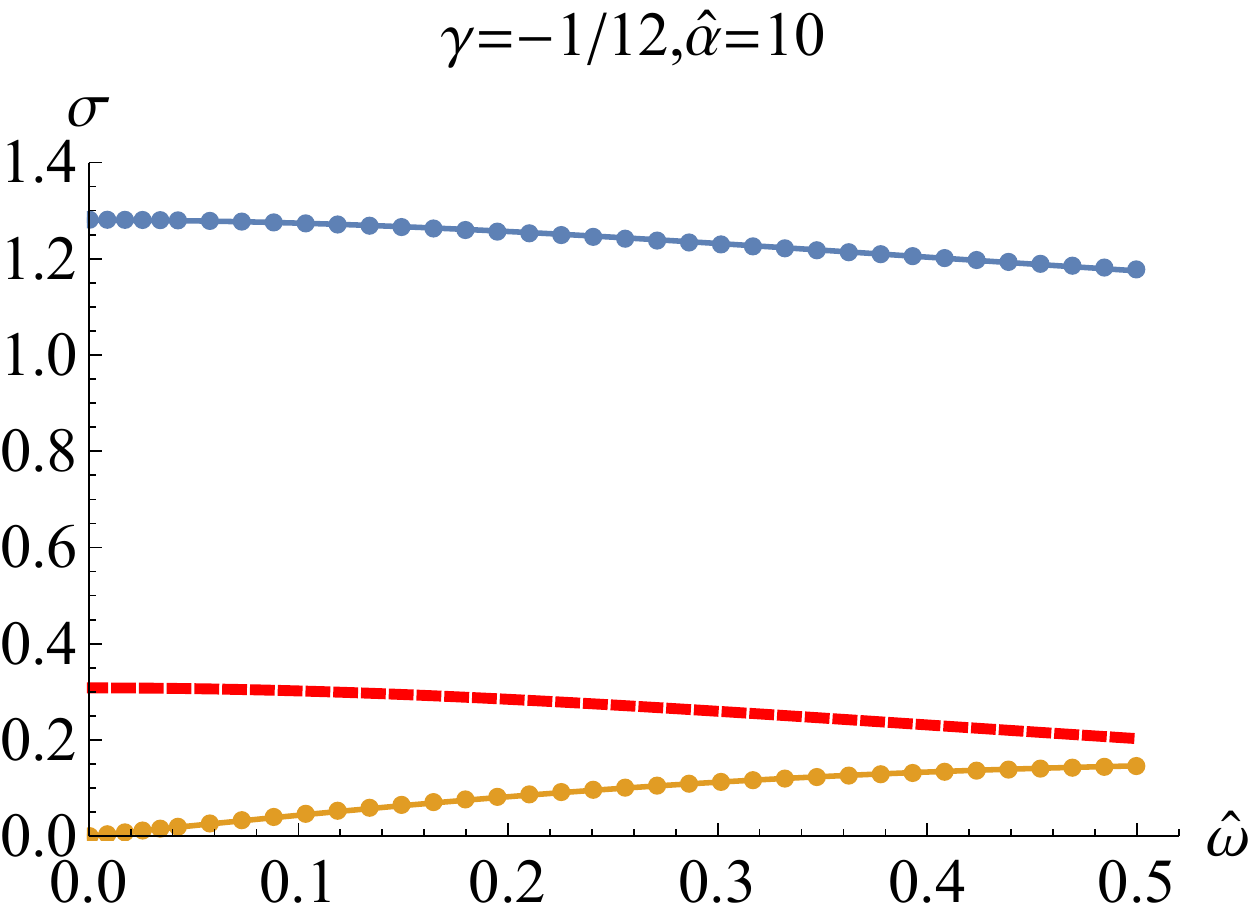}\ \hspace{0.8cm}
\includegraphics[scale=0.6]{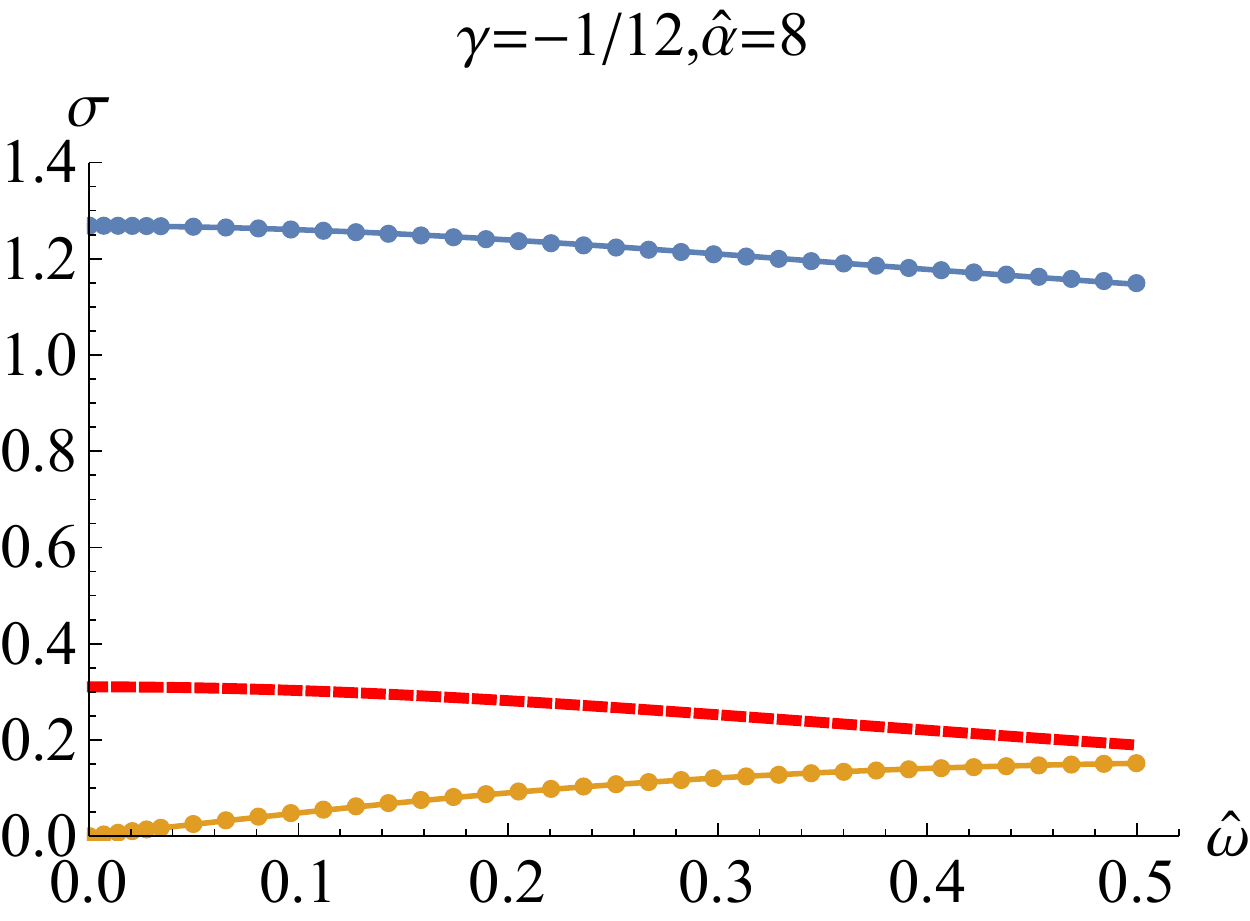}\ \\
\includegraphics[scale=0.6]{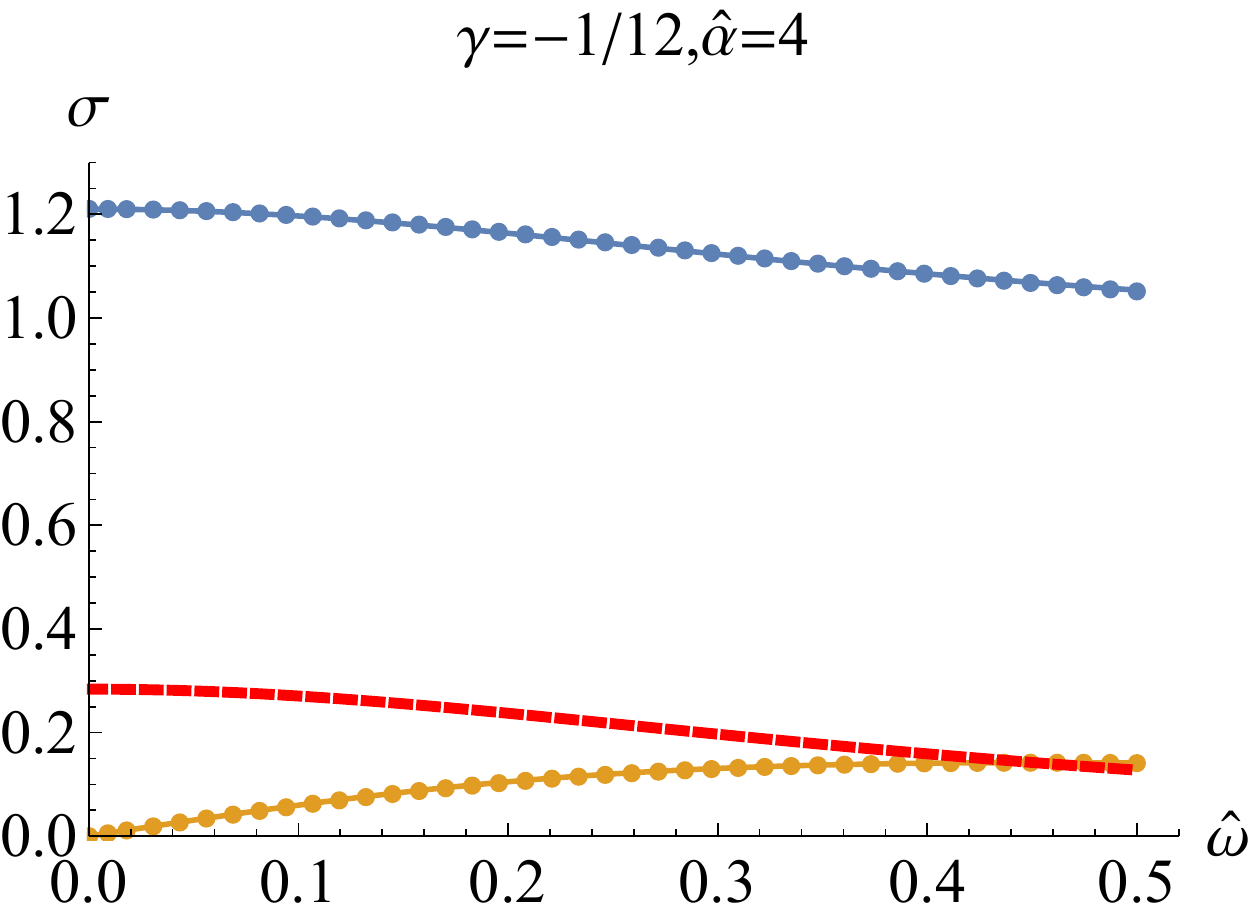}\ \hspace{0.8cm}
\includegraphics[scale=0.6]{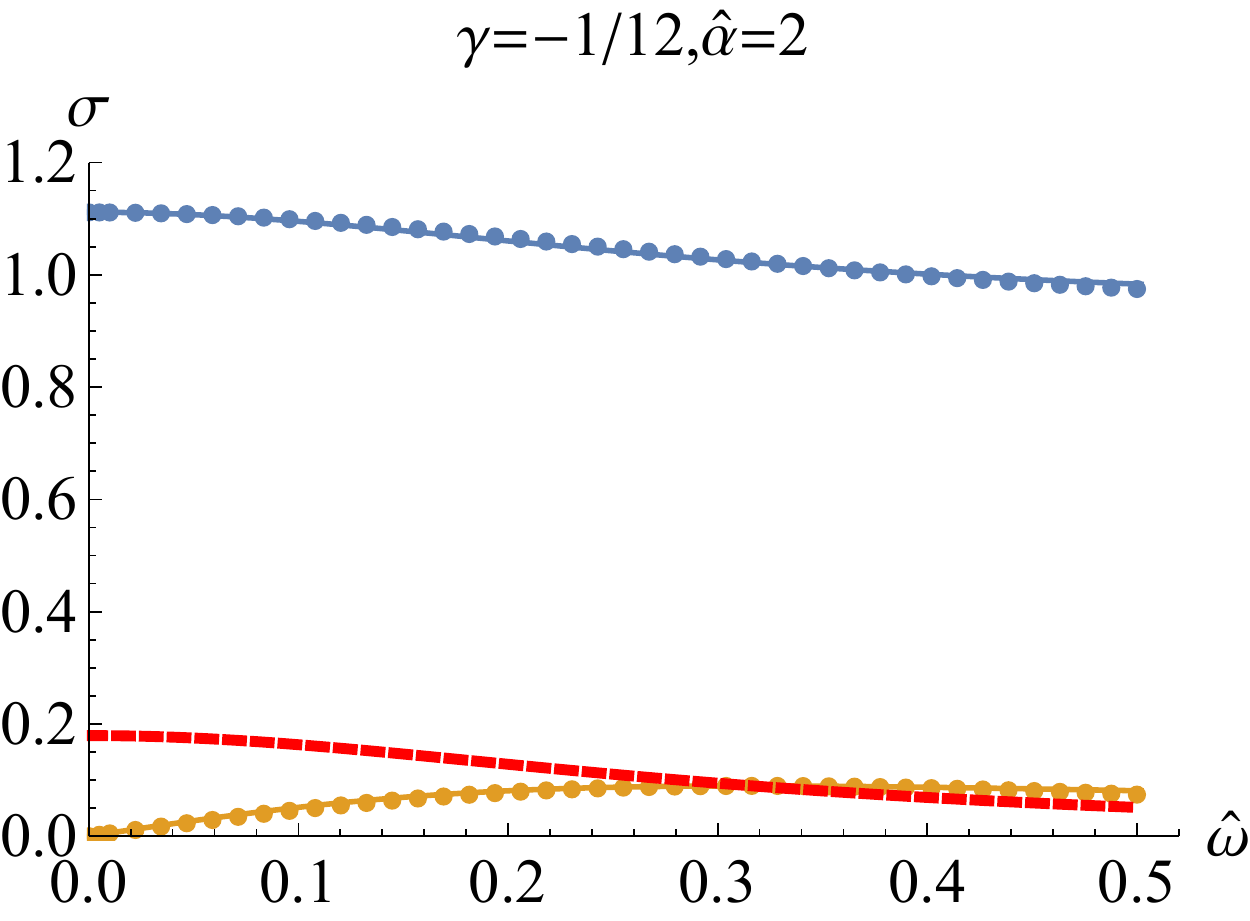}\ \\
\caption{\label{fig-drude_n_gamma_a} The low frequency behavior of the optical conductivity with momentum dissipation for different $\hat{\alpha}$ and $\gamma=-1/12$.
They are fitted by a modified Drude formula (\ref{modified-Drude}).
The red dashed line is the real part of optical conductivity fitted with the standard Drude formula.}}
\end{figure}

We are also interested in the low frequency behavior for $\gamma<0$ and large $\hat{\alpha}$,
in which a peak exhibits (the bottom plots in FIG.\ref{fig-con}).
From FIG.\ref{fig-drude_n_gamma_a}, we see that the non-Drude behavior can be well fitted
with different $\hat{\alpha}$ for $\gamma=-1/12$ by the modified Drude formula (\ref{modified-Drude}).
Quantitatively, we fit $\sigma_Q$ with different $\hat{\alpha}$ for $\gamma=-1/12$ in Table \ref{table-mod-Drude-n}.
Again, it confirms that the modified Drude formula is more suitable to describe the incoherent transport from the momentum dissipation
than that from the Weyl term.

\begin{widetext}
\begin{table}[ht]
\begin{center}
\begin{tabular}{|c|c|c|c|c|c|c|c|}
         \hline
~$\hat{\alpha}$~ &~$10$~&~$8$~&~$6$~&~$4$~&~$2$~
          \\
        \hline
~$\sigma_Q$~ & ~$0.972$~&~$0.957$~&~$0.941$~&~$0.926$~ & ~$0.932$~
          \\
        \hline
\end{tabular}
\caption{\label{table-mod-Drude-n}The characteristic quantity of incoherence $\sigma_Q$ for $\gamma=-1/12$ and different $\hat{\alpha}$.}
\end{center}
\end{table}
\end{widetext}

\section{DC conductivity, diffusion constant and susceptibility}\label{sec-diff-suscept}

In this section, we study the DC conductivity, charge diffusion constant
and susceptibility accommodating with the Maxwell-Wely theory (\ref{ac-ma}) with the momentum dissipation.

There are many ways to calculate these quantities.
Here we shall use the membrane paradigm approach \cite{Kovtun:2003wp,Brigante:2007nu}.
The key point of the membrane paradigm is to define the membrane current on the stretched horizon $u_H=1-\epsilon$ with $\epsilon\ll1$
\fa
j^{\mu}=\frac{1}{4}n_{\nu}X^{\mu\nu\rho\sigma}F_{\rho\sigma}\mid_{u=u_H}\,,
\ffa
where $n_{\nu}$ is a unit radial normal vector.
By the Ohm's law, it is straightforward to write down the expression of the DC conductivity in our present framework \cite{Ritz:2008kh,Myers:2010pk}
\fa
\sigma_0=\sqrt{-g}g^{xx}\sqrt{-g^{tt}g^{uu}X_1X_5}\mid_{u=1}\,.
\label{DC}
\ffa
Moreover, following \cite{Kovtun:2003wp,Brigante:2007nu}, it is also easy to obtain the diffusion constant \cite{Ritz:2008kh,Myers:2010pk}
\fa
D=-\sigma_0\int_0^1\frac{1}{\sqrt{-g}g^{tt}g^{uu}X_3}du\,.
\label{Diffusion}
\ffa
And then, using the Einstein relation $D=\sigma_0/\chi$, the susceptibility can be expressed as \cite{Kovtun:2003wp,Brigante:2007nu}
\fa
\chi^{-1}=-\int_0^1\frac{1}{\sqrt{-g}g^{tt}g^{uu}X_3}du\,.
\ffa
For the details, we can refer to \cite{Ritz:2008kh,Myers:2010pk}.

\begin{figure}
\center{
\includegraphics[scale=0.6]{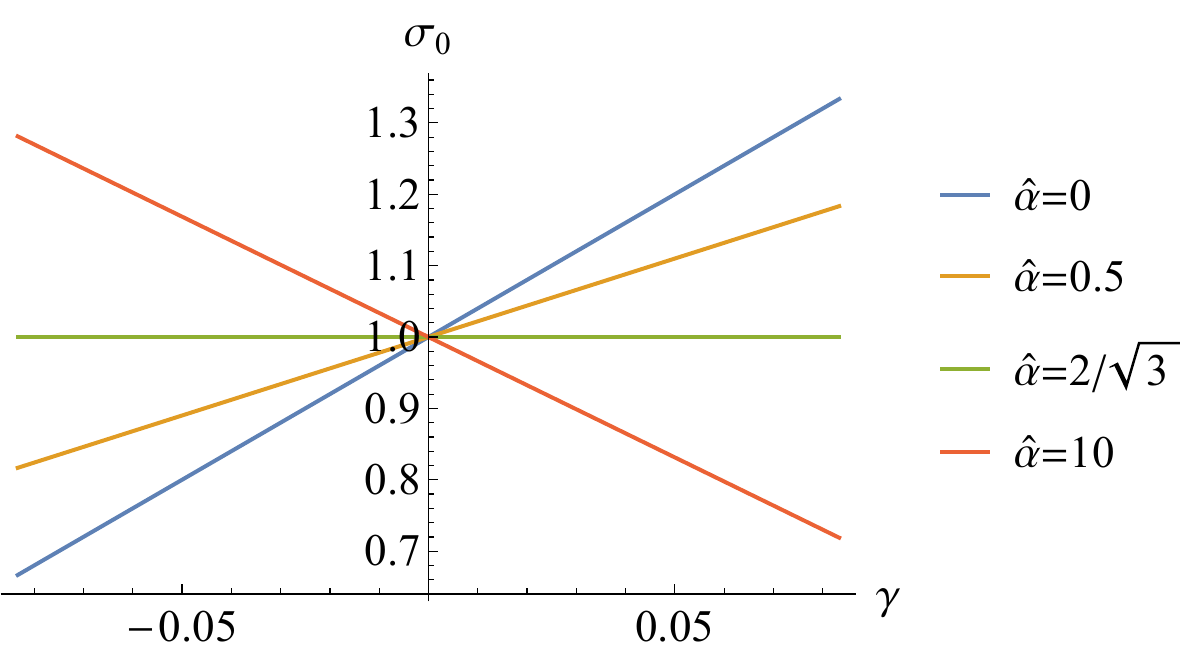}\ \hspace{0.8cm}
\includegraphics[scale=0.6]{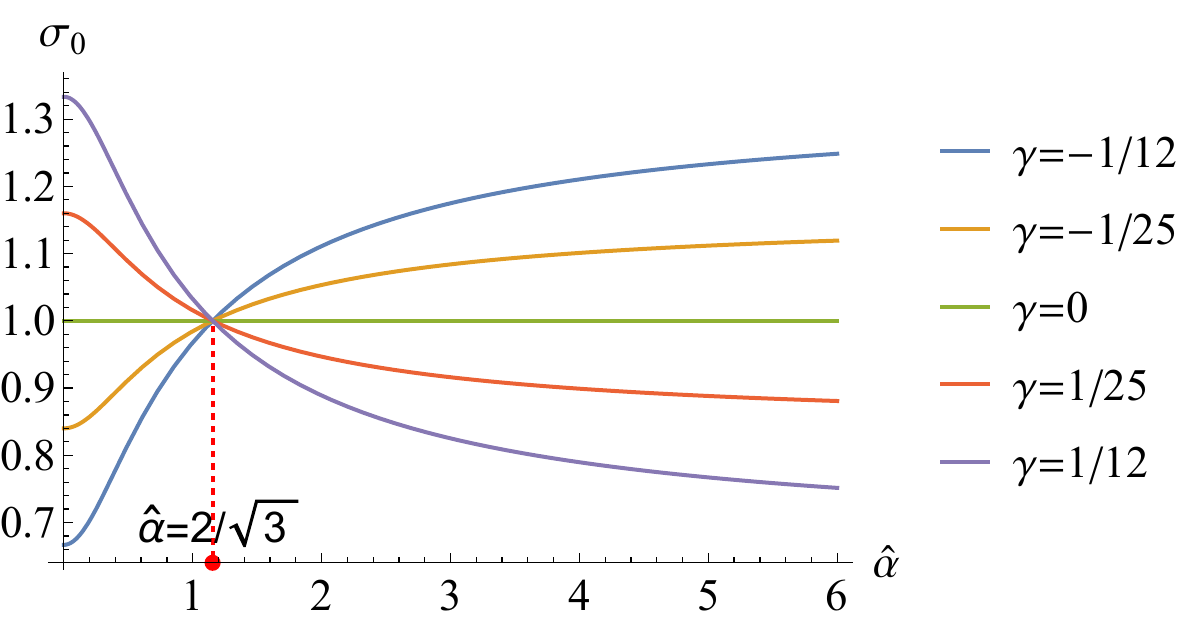}\ \\
\caption{\label{fig-DCvsalpha} Left plot: The DC conductivity $\sigma_0$ versus the Weyl coupling parameter $\gamma$ for some fixed dissipation constant $\hat{\alpha}$.
Right plot: The DC conductivity $\sigma_0$ versus the dissipation constant $\hat{\alpha}$ for some fixed Weyl coupling parameter $\gamma$.}}
\end{figure}

\begin{figure}
\center{
\includegraphics[scale=0.6]{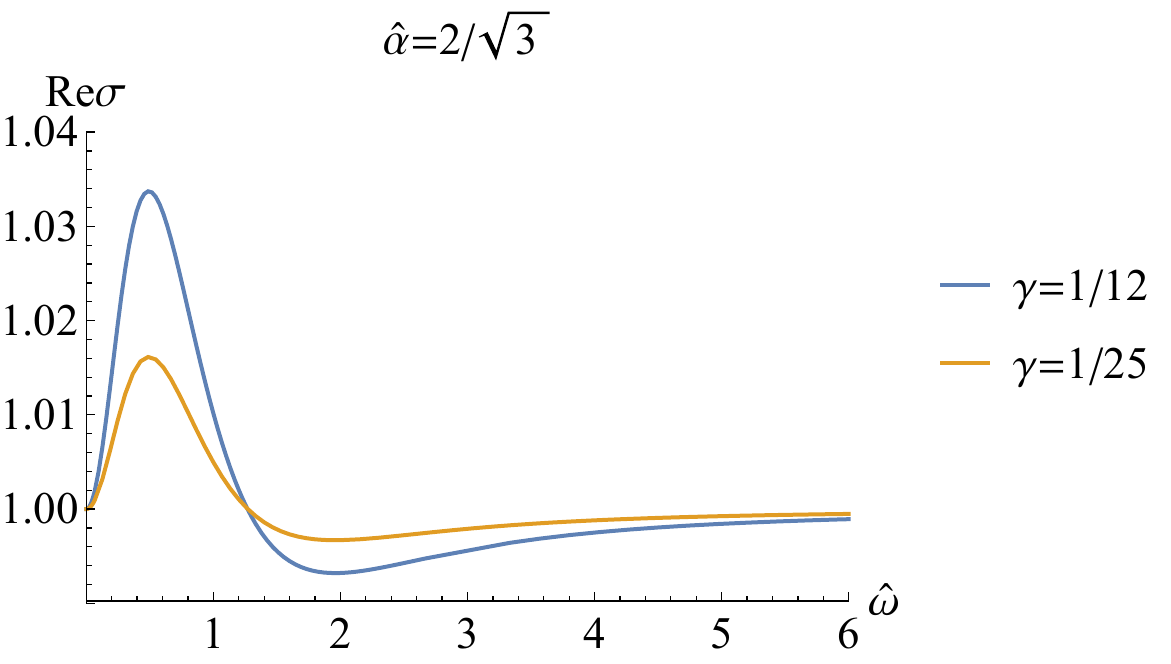}\ \hspace{0.8cm}
\includegraphics[scale=0.6]{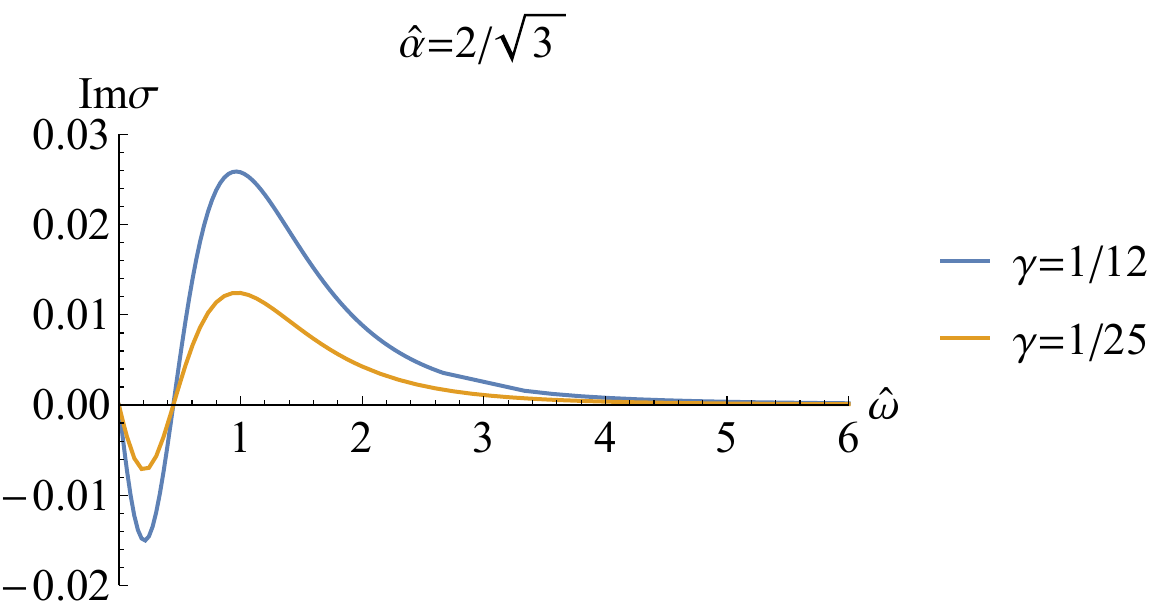}\ \\
\caption{\label{fig-2osqrt3} The real part (left plot) and the imaginary part (right plot) of the optical conductivity as the function of
$\hat{\omega}$ for $\hat{\alpha}=2/\sqrt{3}$ and different values of $\gamma$.
}}
\end{figure}

In the following, we shall explicitly discuss these quantities.
First, evaluating Eq.(\ref{DC}), we obtain the explicit expression of the DC conductivity
\fa
\sigma_0=1+\frac{2}{3}\Big(2+\frac{4(\sqrt{6\hat{\alpha}^2+1}-2\hat{\alpha}^2-1)}{\hat{\alpha}^2}\Big)\gamma\,.
\label{DC-specific}
\ffa
It is obvious that the DC conductivity is linear dependence on $\gamma$ for given $\hat{\alpha}$ (Eq.(\ref{DC-specific}) or see left plot in FIG.\ref{fig-DCvsalpha}).
Also we note that when $\hat{\alpha}=2/\sqrt{3}$, the DC conductivity is independent of the Weyl coupling parameter $\gamma$.
It is a specific point of the conductivity, which shall be further discussed in what follows.
FIG.\ref{fig-DCvsalpha} displays the DC conductivity $\sigma_0$ versus $\gamma$ for different $\hat{\alpha}$ (left plot)
and $\sigma_0$ as the function of $\hat{\alpha}$ for different $\gamma$ (right plot).
Since the momentum dissipation destroys the motion of the particle (vortices), as $\hat{\alpha}$ increases,
the DC conductivity $\sigma_0$ decreases (increases) for $\gamma>0$ ($\gamma<0$).
This picture is consistent with that of the optical conductivity discussed in previous section.
But we would like to point out that though for $\hat{\alpha}=2/\sqrt{3}$, the DC conductivity $\sigma_0=1$,
being independent of the Weyl coupling parameter, the optical conductivity depends on $\gamma$ except for $\omega\rightarrow 0$ and $\infty$ (see FIG.\ref{fig-2osqrt3}).

\begin{figure}
\center{
\includegraphics[scale=0.6]{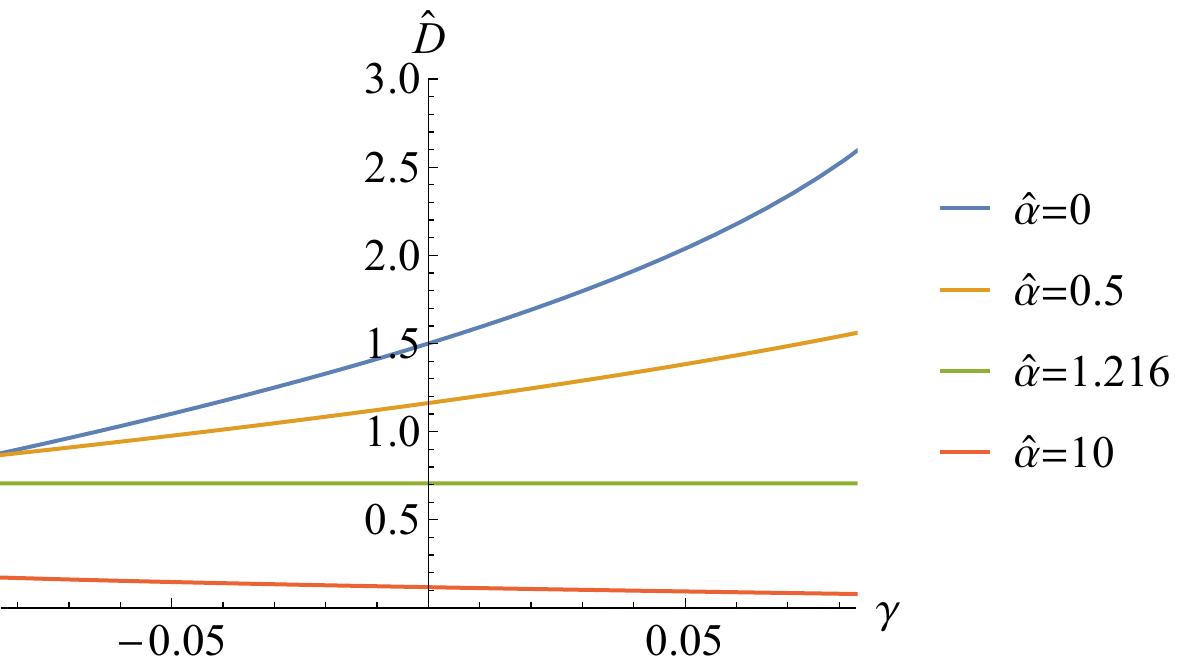}\ \hspace{0.8cm}
\includegraphics[scale=0.6]{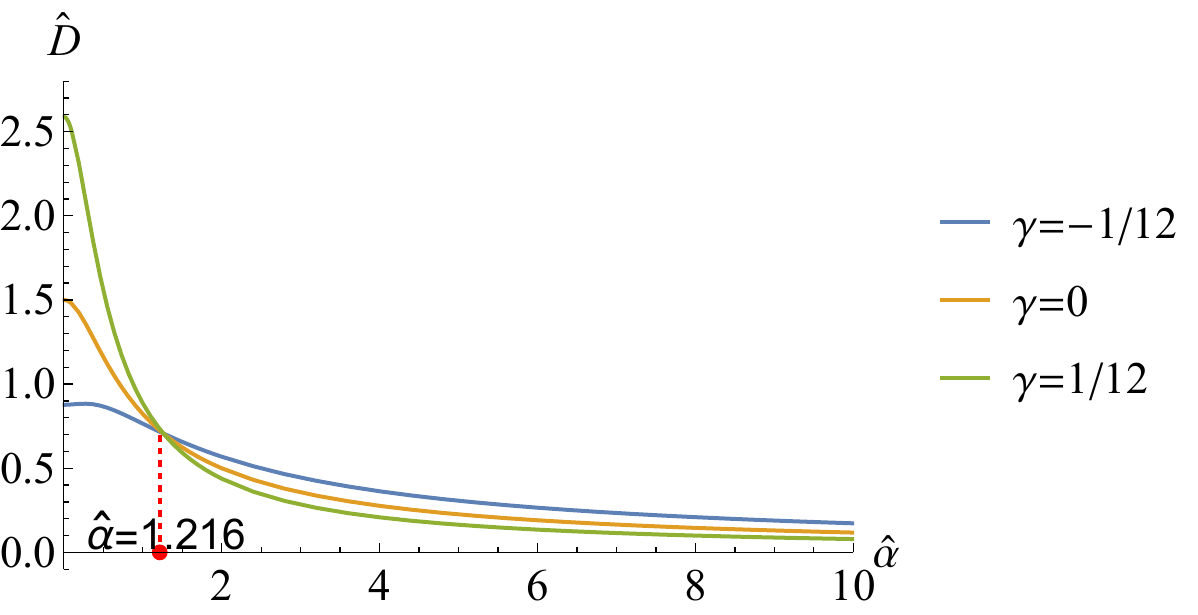}\ \\
\caption{\label{fig-diffusion} Left plot: The charge diffusion constant $\hat{D}$ versus the Weyl coupling parameter $\gamma$ for some fixed dissipation constant $\hat{\alpha}$.
Right plot: The charge diffusion constant $\hat{D}$ versus the dissipation constant $\hat{\alpha}$ for some fixed Weyl coupling parameter $\gamma$.}}
\end{figure}
\begin{figure}
\center{
\includegraphics[scale=0.6]{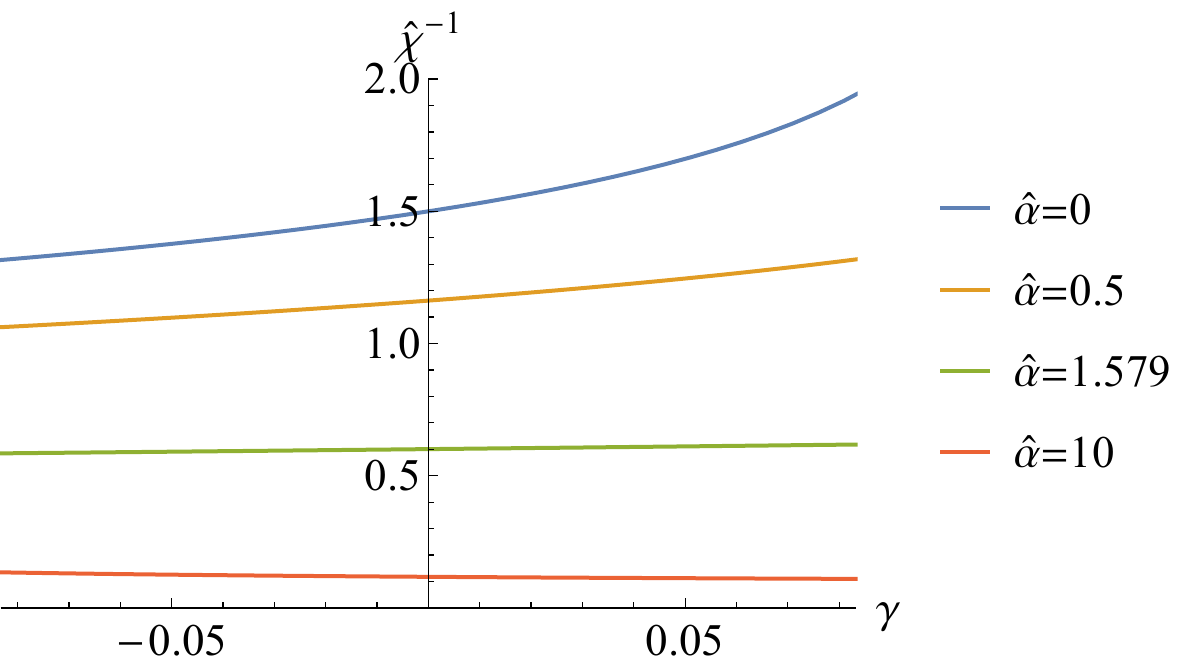}\ \hspace{0.8cm}
\includegraphics[scale=0.6]{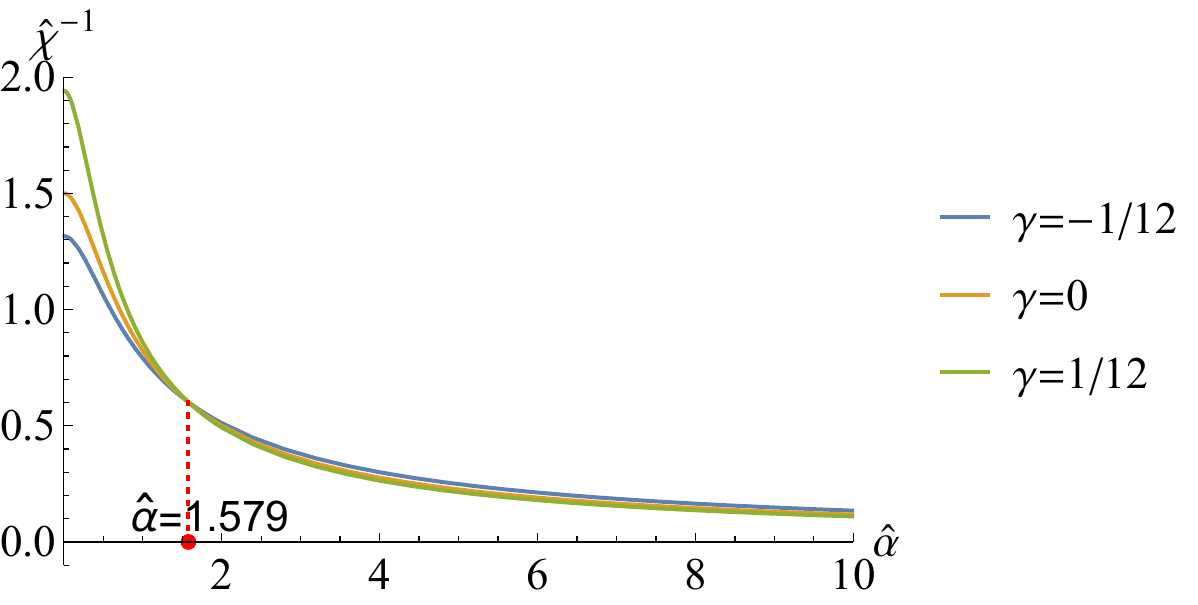}\ \\
\caption{\label{fig-susceptibility} Left plot: The inverse of the susceptibility $\hat{\chi}^{-1}$ versus the Weyl coupling parameter $\gamma$ for some fixed dissipation constant $\hat{\alpha}$.
Right plot: The inverse of the susceptibility $\hat{\chi}^{-1}$ versus the dissipation constant $\hat{\alpha}$ for some fixed Weyl coupling parameter $\gamma$.}}
\end{figure}

The diffusion constant and the susceptibility with Weyl correction in the boundary field theory dual to
the SS-AdS geometry can be analytically worked out \cite{Myers:2010pk,Ritz:2008kh}.
However, for the EA-AdS geometry (\ref{bl-br}) and (\ref{fu}), it is difficult to analytically derive their expressions and we need to resort to the numerical method.
We have exhibited the dimensionless charge diffusion constant $\hat{D}\equiv 2\pi T D$
and the inverse of the dimensionless susceptibility $\hat{\chi}^{-1}$ with $\hat{\chi}\equiv 2\pi T \chi$
as the function of the coupling $\gamma$ for some fixed $\hat{\alpha}$ in the left plots in FIG.\ref{fig-diffusion}
and FIG.\ref{fig-susceptibility}, respectively.
While the right plots in FIG.\ref{fig-diffusion} and FIG.\ref{fig-susceptibility}
shows $\hat{D}$ and $\hat{\chi}^{-1}$ versus $\hat{\alpha}$ for different $\gamma$, respectively.
We summarize the momentum dissipation effect on the diffusion constant and the susceptibility as follows.
First, similar to the case of DC conductivity, there is a specific value
$\hat{\alpha}\approx1.216$ for $\hat{D}$ and $\hat{\alpha}\approx1.578$ for $\hat{\chi}^{-1}$,
for which the diffusion constant $\hat{D}$ and the susceptibility $\hat{\chi}$ are independent of the Weyl coupling parameter $\gamma$.
We note that for different observables, $\sigma_0$, $\hat{D}$ and $\hat{\chi}^{-1}$,
the specific values of $\hat{\alpha}$ is different and so these values are not universal.
Second, when $\hat{\alpha}<1.216$, the diffusion constant increases with the increase of $\gamma$.
But the tendency of the increase of the $\hat{D}$ as the function $\gamma$ tends to slow down with the increase of $\hat{\alpha}$.
While for $\hat{\alpha}>1.216$, the case is opposite, \emph{i.e.},
with the increase of $\gamma$, $\hat{D}$ decreases.
But the tendency of the decrease is weak.
Similar phenomena can be found for the susceptibility.
Third, the momentum dissipation suppresses the diffusion constant and the inverse of the susceptibility,
regardless of the sign of $\gamma$.
It implies that the momentum dissipation has similar effect on the diffusion constant or the susceptibility
regardless of the excitation being particles or vortices.
We shall further understand and explore such phenomenon and their microscopic mechanism in future.

\section{EM duality}\label{sec-EM-duality}

In Section \ref{sec-HF} and Appendix \ref{Bounds} (also see \cite{Myers:2010pk}), it has been illustrated that for the metric of the background geometry
being diagonal, the EM duality holds and for very small $\gamma$, the original EM theory relates its dual theory by changing the sign of $\gamma$.
In the AdS/CFT correspondence, the bulk EM duality corresponds to the particle-vortex duality
in the boundary field theory, in which the optical conductivity of the dual theory
is the inverse of that of its original theory \cite{Myers:2010pk,WitczakKrempa:2012gn}
\fa
\sigma_{\ast}(\hat{\omega};\hat{\alpha},\gamma)=\frac{1}{\sigma(\hat{\omega};\hat{\alpha},\gamma)}\,.
\label{sigma-B-A}
\ffa
The proof can be found in \cite{Myers:2010pk,WitczakKrempa:2012gn}.
And we also refer to \cite{Herzog:2007ij} for the derivation of the above relation in a specific class of CFTs.
Since for small $\gamma$, the change of the sign of $\gamma$ corresponds to an approximate particle-vortex duality effect,
we have \cite{Myers:2010pk,WitczakKrempa:2012gn,WitczakKrempa:2013ht,Witczak-Krempa:2013nua,Witczak-Krempa:2013aea,Katz:2014rla}
\fa
\sigma(\hat{\omega};\hat{\alpha},\gamma)\approx\frac{1}{\sigma(\hat{\omega};\hat{\alpha},-\gamma)}\,,~~~~~~~|\gamma|\ll1\,.
\label{sigma-B-A-gamma}
\ffa
From Eqs.(\ref{sigma-B-A}) and (\ref{sigma-B-A-gamma}), we can conclude the following relation
\fa
\sigma_{\ast}(\hat{\omega};\hat{\alpha},\gamma)\approx\sigma(\hat{\omega};\hat{\alpha},-\gamma)
\,,~~~~~~~|\gamma|\ll1\,.
\label{sigma-v1}
\ffa
The above equation indicates that the optical conductivity of the dual EM theory
is approximately equal to that of its original theory for the oppositive sign of $\gamma$.
It also have been explicitly illustrated for $\gamma=\pm 1/12$ in Figure 5 in \cite{Myers:2010pk},
from which we can obviously see that the conductivity of the dual EM theory
is not precisely equal to that of its original theory for the oppositive sign of $\gamma$ except for $\hat{\omega}\rightarrow\infty$.
Next, we shall explore the effect of the momentum dissipation on the EM duality by explicitly presenting
the frequency dependent conductivity of the original theory and its dual theory.

\begin{figure}
\center{
\includegraphics[scale=0.55]{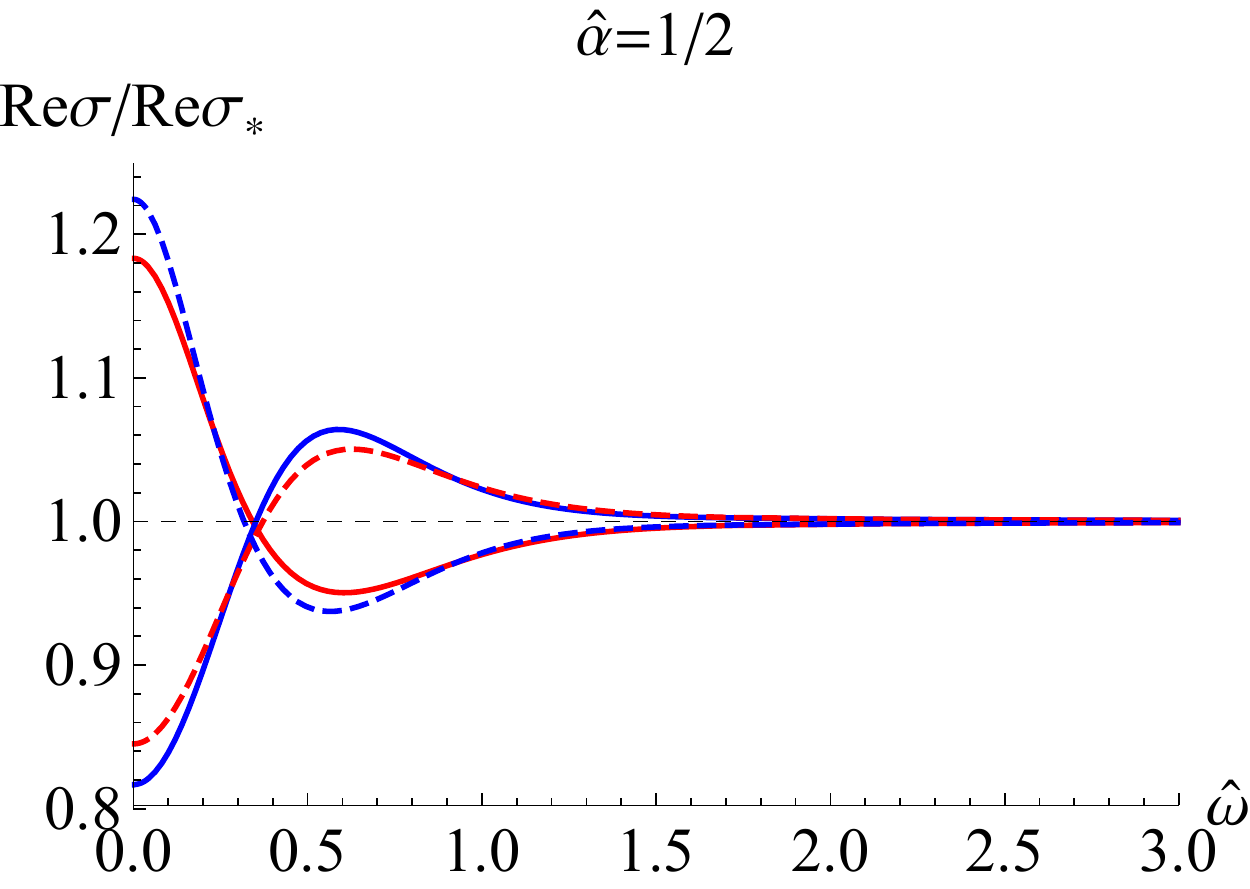}\ \hspace{0.8cm}
\includegraphics[scale=0.55]{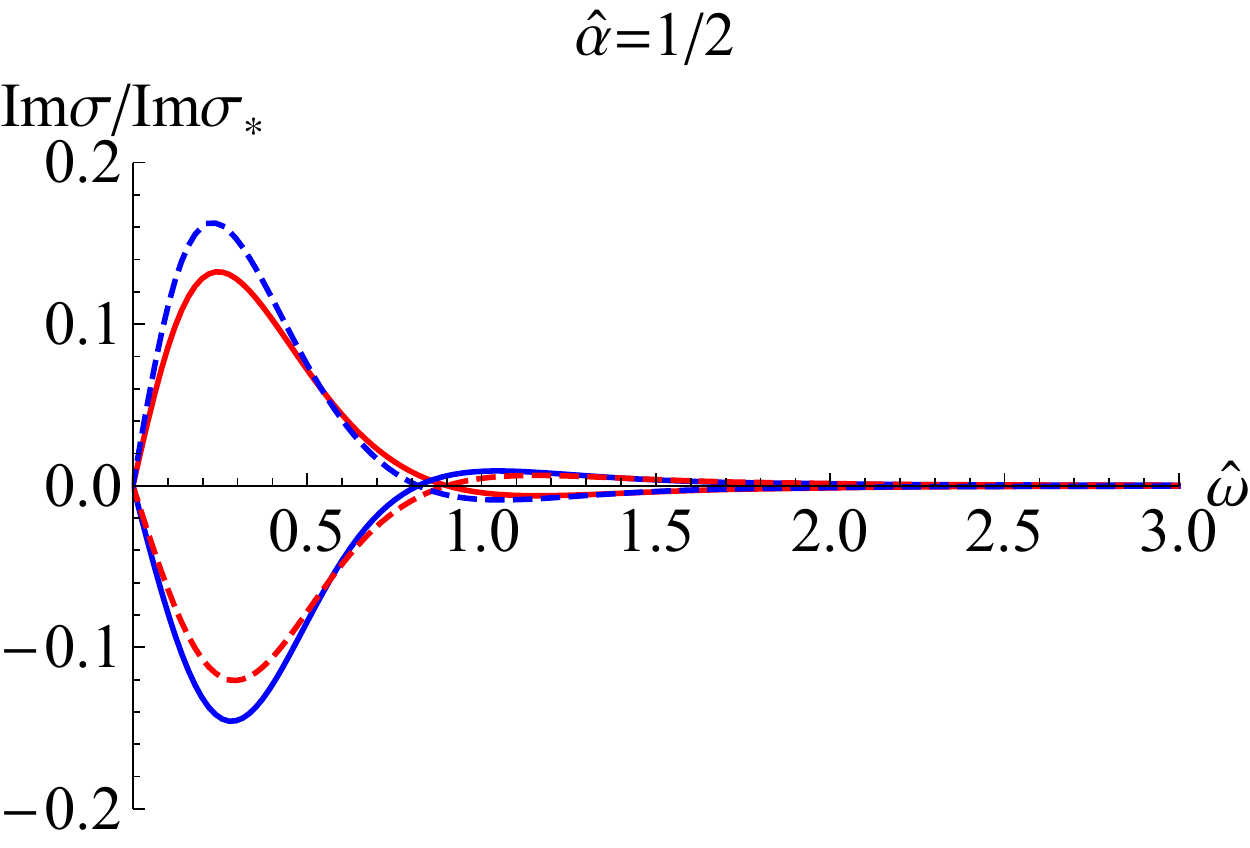}\ \\
\includegraphics[scale=0.55]{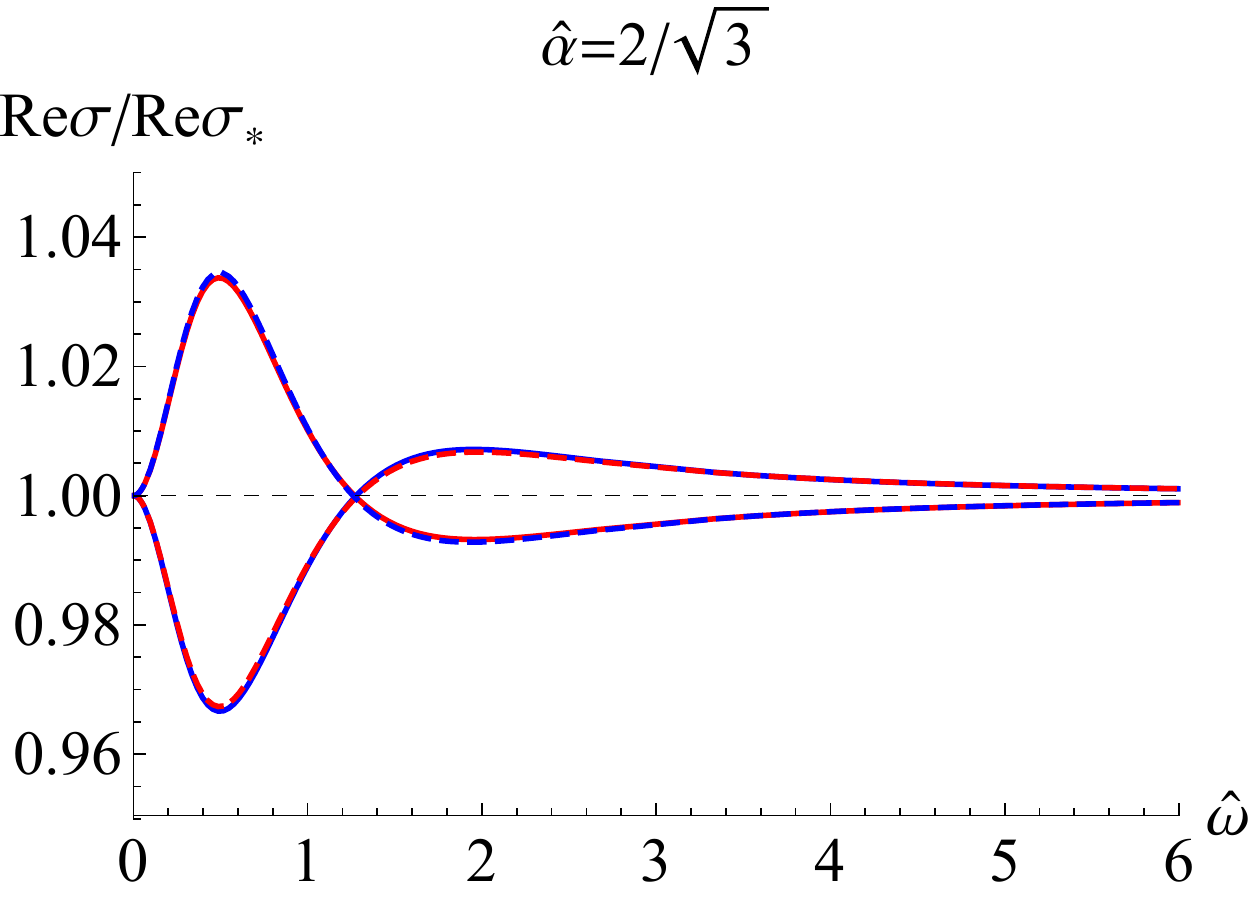}\ \hspace{0.8cm}
\includegraphics[scale=0.55]{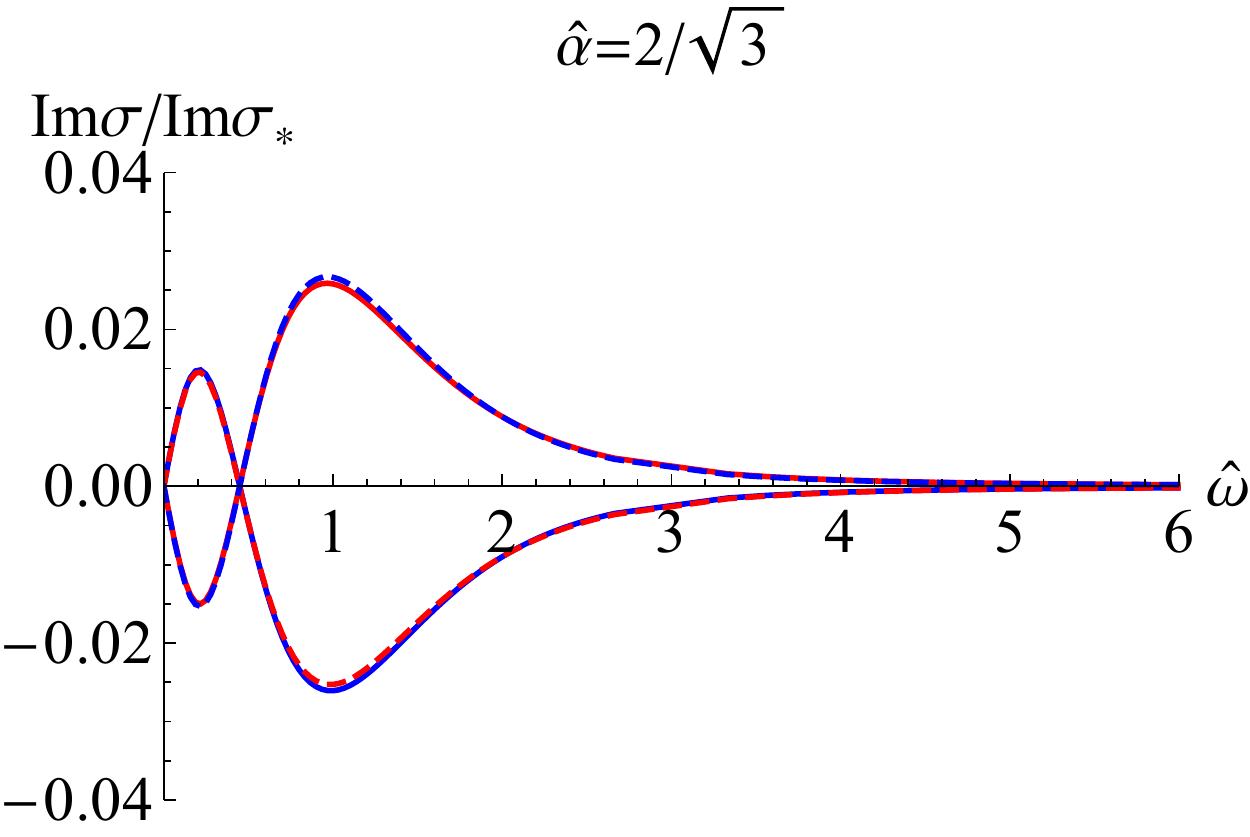}\ \\
\includegraphics[scale=0.55]{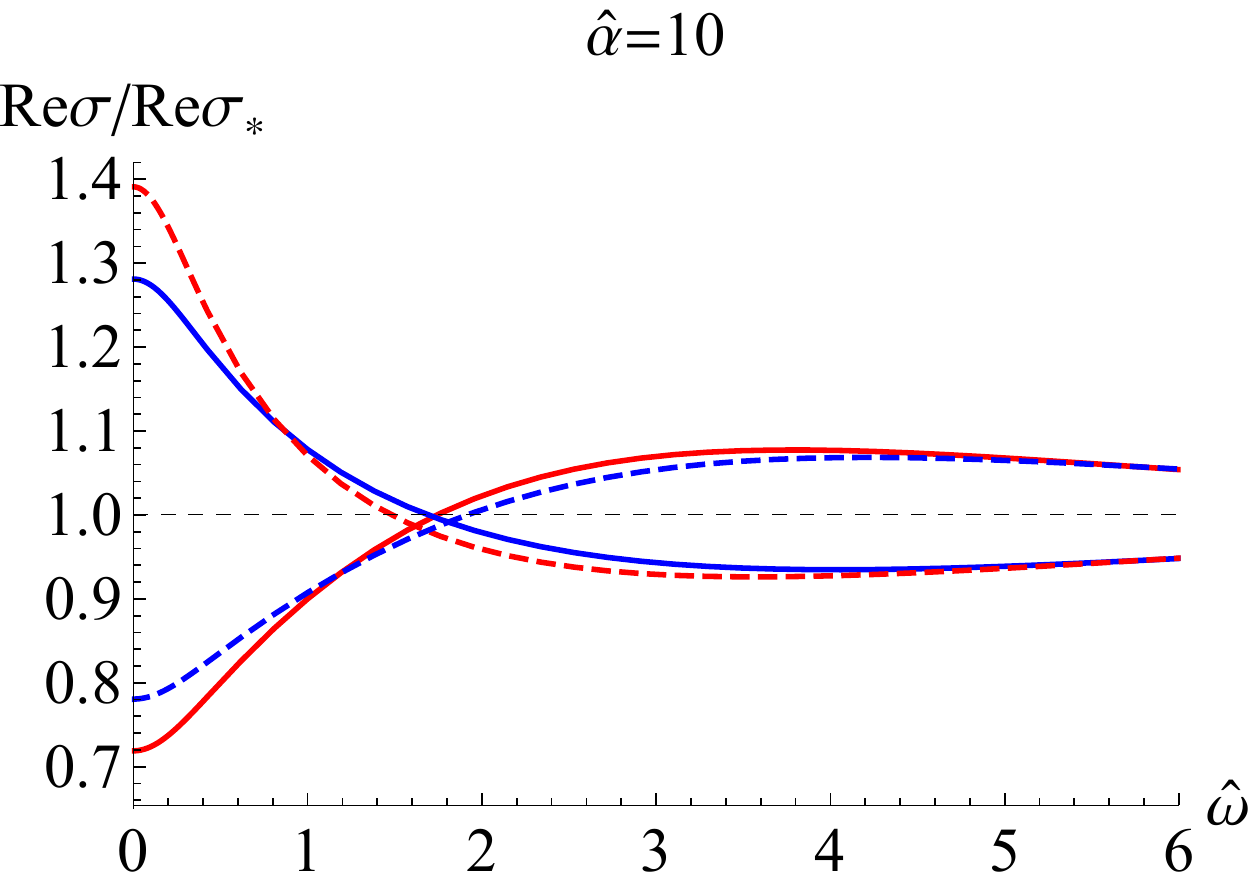}\ \hspace{0.8cm}
\includegraphics[scale=0.55]{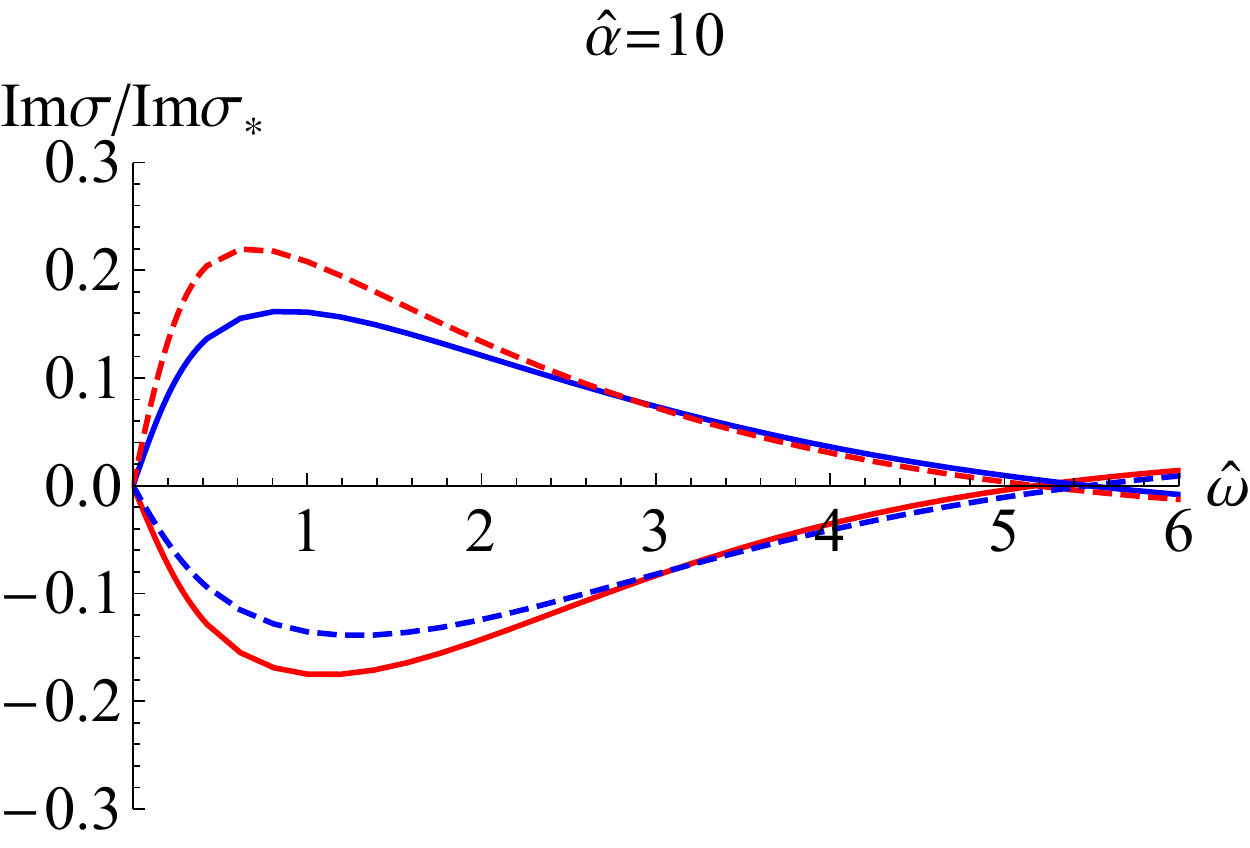}\ \\
\caption{\label{fig-dual} The real part (left plot) and the imaginary part (right plot) of the optical conductivity as the function of
$\hat{\omega}$ for various values of $\gamma$ and $\hat{\alpha}$.
The solid curves are the conductivity of the original EM theory (\ref{ac-SA}),
which have been shown previously in FIG.\ref{fig-con} (red for $\gamma=1/12$ and blue for $\gamma=-1/12$).
While the dashed curves display the conductivity of the EM dual theory (\ref{ac-SB}) for the same value of $\gamma$ and $\hat{\alpha}$.
}}
\end{figure}

First, we focus on the DC conductivity, which can be analytically derived (see Eq.(\ref{DC-specific})).
To this end, we also derive the DC conductivity of the dual EM theory (\ref{ac-SB}) as
\fa
\sigma_{\ast0}(\hat{\alpha},\gamma)&=&\frac{1}{1+\frac{2}{3}\Big(2+\frac{4(\sqrt{6\hat{\alpha}^2+1}-2\hat{\alpha}^2-1)}{\hat{\alpha}^2}\Big)\gamma}
\nonumber
\\
&\approx&1-\frac{2}{3}\Big(2+\frac{4(\sqrt{6\hat{\alpha}^2+1}-2\hat{\alpha}^2-1)}{\hat{\alpha}^2}\Big)\gamma\,.
\label{DC-specific-dual}
\ffa
The second line ($\approx$) is for $|\gamma|\ll1$.
Similarly with $\sigma_0$, we have a specific value $\hat{\alpha}=2/\sqrt{3}$, for which
$\sigma_{\ast0}=1$ and is independent of $\gamma$.
Specifically, at this value of $\hat{\alpha}$, we have $\sigma_{\ast0}=\sigma_0=1$.

Subsequently, we turn to study the optical conductivity.
FIG.\ref{fig-dual} displays the real and imaginary part of the optical conductivity of the bulk EM theory
and its dual EM theory as the function of the frequency $\hat{\omega}$ for $\gamma=\pm1/12$ and various values of $\hat{\alpha}$.
As expected, an oppositive picture appears in the dual EM theory.
That is to say, for small $\hat{\alpha}$, a peak at small frequency in optical conductivity occurs for $\gamma=-1/12$
while a dip exhibits for $\gamma=1/12$ and for large $\hat{\alpha}$, the case is opposite.
In addition, FIG.\ref{fig-dual} also further illustrates that the relation (\ref{sigma-B-A-gamma}) or (\ref{sigma-v1}) holds only for $|\gamma|\ll1$.
But for the specific value of $\hat{\alpha}=2/\sqrt{3}$, which is previously found in DC conductivity,
the optical conductivities of the original EM theory and its dual EM theory
are almost exactly inverses of each other for either value of $\gamma$, \emph{i.e.},
the relation (\ref{sigma-v1}) holds very well.
It indicates that the particle-vortex duality exactly holds for $\hat{\alpha}=2/\sqrt{3}$.
Note that our result is numerically worked out. The analytical derivation and understanding deserves pursuing in future.

Finally, we would like to point out that the specific point $\hat{\alpha}=2/\sqrt{3}$
is not the self-duality point of our present model because we don't have $\hat{X}=X$
at this point such that the dual theory (\ref{ac-SB}) is not identical with it original theory (\ref{ac-SA}).
Therefore, this point is different from the point found in the thermal conductivity of the Maxwell theory in EA-AdS geometry,
at which point the thermal conductivity is independent of the frequency \cite{Davison:2014lua}.

\section{Discussions and open questions}\label{sec-discussion}

In this paper, we have studied the transports, in particular the electric conductivity, in a neutral plasma with momentum dissipation
dual to the Maxwell-Weyl system in EA-AdS geometry.
In previous studies \cite{Myers:2010pk,WitczakKrempa:2012gn,WitczakKrempa:2013ht,Witczak-Krempa:2013nua,Witczak-Krempa:2013aea,Katz:2014rla},
they find that the optical conductivity in the neutral system at finite temperature dual to the Maxwell-Weyl system in SS-AdS geometry
exhibits a peak or a dip depending on the sign of the Weyl coupling parameter $\gamma$.
It provides a possible description of the CFT of the boson Hubbard model in holographic framework \cite{Sachdev:2011wg}.
But there is still a degree of freedom of the sign of $\gamma$.
Our main results for the optical conductivity without self-duality but with the momentum dissipation are displayed in FIG.\ref{fig-con}
and the corresponding physical interpretation is presented in Section \ref{sec-optical-conductivity}.
For $\gamma>0$, the strong disorder drive the peak in the low frequency optical conductivity
into a dip. While for $\gamma<0$, an oppositive scenario happens. That is to say,
the dip in optical conductivity at low frequency
gradually upgrades and eventually develops into a peak with the increase of the disorder.
Our present model provides a route toward the problem that which sign of $\gamma$ is the correct description of the CFT of boson Hubbard model.
Also we have quantitatively studied the low frequency behavior of the optical conductivity
by using the modified Drude formula (\ref{modified-Drude}) to fit our numerical data.
It provides a hint regarding the coherent or incoherent contribution from the Weyl term and the momentum dissipation
and deserves further studying.
Further, we find that there is a specific value of the momentum dissipation constant $\hat{\alpha}=2/\sqrt{3}$,
for which the DC conductivity $\sigma_0$ is independent of $\gamma$ and the particle-vortex duality related by the change of the sign of $\gamma$ holds very well.
Aside from the conductivity, there is also a specific value of $\hat{\alpha}$ for the diffusion constant and susceptibility,
for which these quantities are independent of the Weyl coupling parameter $\gamma$.
But these specific values of $\hat{\alpha}$ are different from each other
and so they are not universal in this Maxwell-Weyl system in EA-AdS geometry.

In addition, we also present several comments on the physics of our present model as follows.
References \cite{Myers:2010pk,WitczakKrempa:2012gn,WitczakKrempa:2013ht,Witczak-Krempa:2013nua,Witczak-Krempa:2013aea,Katz:2014rla}
introduce an extra Weyl tensor $C_{\mu\nu\rho\sigma}$ coupling to Maxwell field
in the SS-AdS black brane to study the QC transports in the neutral plasma at finite temperature.
Another way to study the QC physics in holographic framework is
that by incorporating a neutral bulk scalar field interacting with gravity,
the corresponding scalar operator has an expected value \cite{Myers:2016wsu}.
Setting the source of scalar field in the dual boundary theory to zero,
we can study the QC physics \cite{Myers:2016wsu}.
If we set the source of scalar field in the dual boundary field theory nonvanishing,
the model \cite{Myers:2016wsu} also provides a starting point to study the physics away from QCP.
In our present model, $\Phi_I$ correspond to turning on sources being spatial linear in the dual boundary theory,
which introduces a dimensionful parameter $\alpha$ into the dual boundary theory
and so the physics we studied is that away from QCP.
Therefore, our results also provide some insight into the proximity effect in QCP.
More comprehensive analysis on this point will be discussed elsewhere.

Finally, we comment some open questions deserving further exploration.
\begin{enumerate}
\item In this paper, we mainly study the optical conductivity at the zero momentum,
which is relatively easy to calculate since the equations of motion simplify to a great extent.
But the transports at the finite momentum and energy in condensed matter laboratories
have been obtained now or shall be given in near future \cite{Zaanen:2015,Ammon:2015,Keimer:2015,Forcella:2014gca,Vig:2015},
which reveal more information of the systems.
On the other hand, in \cite{WitczakKrempa:2013ht} they have also studied the responses of Maxwell-Weyl system in SS-AdS geometry at the finite momentum
and found that it indeed provided far deeper insights into this system than that at the zero momentum.
Therefore it is interesting and important to further study the responses of our present model with the momentum dissipation at full momentum and energy spaces.
\item The spatial linear dependent axionic fields are the simplest way to implement the momentum dissipation, or say disorder.
We can also introduce the momentum dissipation by incorporating the higher order terms of axions \cite{Baggioli:2014roa,Alberte:2015isw,Alberte:2016xja,Baggioli:2016oqk,Gouteraux:2016wxj}
to study the properties of transport of the Maxwell-Weyl system. The higher order terms of axions induce metal-insulator transition (MIT) \cite{Baggioli:2014roa,Baggioli:2016oqk}
and provide a way to study the properties of solid in the holographic framework \cite{Alberte:2015isw,Alberte:2016xja}.
Also an insulating ground state can be obtained in this way \cite{Gouteraux:2016wxj}.
In addition, another mechanism of momentum dissipation in holographic framework is the Q-lattice \cite{Donos:2013eha,Donos:2014uba},
by which various type of holographic MIT model have been built \cite{Donos:2013eha,Donos:2014uba,Ling:2015epa,Ling:2015exa}.
It is certainly interesting and valuable to incorporate Q-lattice responsible for the momentum dissipation into the Maxwell-Weyl system
and study its transport behavior, in particular to see how universal our results presented in this paper are.
\item Another important transport quantity is the magneto-transport, 
which has been studied when the momentum dissipation is presented in \cite{Amoretti:2014zha,Amoretti:2014mma,Amoretti:2015gna,Amoretti:2016cad}.
It is interesting to study the magneto-transport property in our framework
and explore the meaning of EM duality.
\item We would like to study the holographic superconductor in our present framework.
In \cite{Wu:2010vr}, the holographic superconductor with Weyl term is constructed.
A main result is that the ratio of the gap frequency over the superconducting critical temperature $\omega_g/T_c$
runs with the Weyl parameter $\gamma$. Subsequently,
a series of works study such holographic superconducting systems with Weyl term,
see for example \cite{Ma:2011zze,Momeni:2011ca,Momeni:2012ab,Zhao:2012kp,Momeni:2013fma,Momeni:2014efa,Zhang:2015eea,Mansoori:2016zbp}.
It would be interesting and useful to study the holographic superconducting systems without self-duality but with the momentum dissipation
and further reveal the role that the momentum dissipation play in the Maxwell-Weyl system.
\item A challenging question is to obtain a full backreaction solution for the EMA-Weyl system.
As has been pointed out in \cite{Myers:2010pk,Ling:2016dck}, we need to develop new numerical technics to solve differential equations beyond the second order with high
nonlinearity.
\end{enumerate}
We plan to explore these questions and publish our results in the near future.

\begin{acknowledgments}

We are very grateful to R. C. Myers, and W. Witczak-Krempa for comments on the QCP.
We are also very grateful to Peng Liu, A. Lucas and W. Witczak-Krempa
for many useful discussions and comments on the manuscript.
We are also grateful to Yi Ling, Peng Liu and Zhenhua Zhou for the collaboration in the related projects.
This work is supported by the Natural Science Foundation of China under
Grant Nos.11305018, 11275208, by the Program for Liaoning Excellent Talents in University (No.LJQ2014123),
and by the grant (No.14DZ2260700) from the Opening Project of Shanghai Key Laboratory
of High Temperature Superconductors.

\end{acknowledgments}

\begin{appendix}

\section{Bounds on the coupling}\label{Bounds}

In \cite{Myers:2010pk,Ritz:2008kh}, they examine the causality of the boundary CFT dual to the 4-dimensional SS-AdS geometry
and the instabilities of the vector modes in this holographic model and find that
a constraint should be imposed on the coupling $\gamma$ as $\gamma\in\mathcal{S}_{0}$ with $\mathcal{S}_0:=[-1/12,1/12]$.
In this appendix, we examine the constraint on $\gamma$ in our present holographic model following \cite{Myers:2010pk,Ritz:2008kh}.

Now, we turn on the perturbations of the gauge field and decompose it in the Fourier space as
\fa
\label{A-Fourier}
A_{\mu}(t,x,y,u)=\int\frac{d^3q}{(2\pi)^3}e^{i\bf{q}\cdot\bf{x}}A_{\mu}(u,\bf{q})
\,,
\ffa
where $\textbf{q}\cdot\textbf{x}=-\omega t+ q^x x + q^y y$.
Without loss of generality, we set $\textbf{q}^{\mu}=(\omega,q,0)$ and choose the gauge fixed as $A_{u}(u,\textbf{q})=0$.
Further, it is convenient to write the tensor $X_{\mu\nu}^{\ \ \rho\sigma}$ as \cite{Myers:2010pk}
\fa
X_{A}^{\ B}=\{X_1(u),X_2(u),X_3(u),X_4(u),X_5(u),X_6(u)\}\,,
\label{XAB}
\ffa
with
\fa
A,B\in\{tx,ty,tu,xy,xu,yu\}\,.
\ffa
When the background is rotationally symmetric in $xy$-plane, one has $X_1(u)=X_2(u)$ and $X_5(u)=X_6(u)$.
Note that from Eq. (\ref{X-hat}), one can easily find that $\widehat{X}_A^{\ B}$ is also diagonal and its entries $\widehat{X}_i$ is the inverse of $X_i$, \emph{i.e.}, $\widehat{X}_i=1/X_i$.

Given $X$, the equations of motion (\ref{eom-Max}) in the planar black brane geometry (\ref{bl-br}) can be evaluated as \cite{Myers:2010pk}
\fa
&&
A'_t+\frac{\hat{q}f}{\hat{\omega}}\frac{X_5}{X_3}A'_x=0\,,
\label{Ma-AtI}
\\
&&
A''_t+\frac{X'_3}{X_3}A'_t
-\frac{\mathfrak{p}^2\hat{q}}{f}\frac{X_1}{X_3}\big(\hat{q}A_t+\hat{\omega}A_x\big)
=0\,,
\label{Ma-AtII}
\\
&&
A''_x+\Big(\frac{f'}{f}
+\frac{X'_5}{X_5}\Big)A'_x
+\frac{\mathfrak{p}^2\hat{\omega}}{f^2}\frac{X_1}{X_5}\big(\hat{q}A_t+\hat{\omega}A_x\big)=0\,,
\label{Ma-Ax}
\\
&&
A''_y
+\Big(\frac{f'}{f}+\frac{X'_6}{X_6}\Big)A'_y
+\frac{\mathfrak{p}^2}{f^2}\Big(\hat{\omega}^2\frac{X_2}{X_6}-\hat{q}^2f\frac{X_4}{X_6}\Big)A_y
=0\,,
\label{Ma-Ay}
\ffa
where the prime denotes the derivative with respect to $u$ and
we have defined the dimensionless frequency and momentum as
\fa
\hat{\omega}\equiv\frac{\omega}{4\pi T}=\frac{\omega}{\mathfrak{p}}\,,
~~~~~
\hat{q}\equiv\frac{q}{4\pi T}=\frac{q}{\mathfrak{p}}\,,
~~~~~
\mathfrak{p}\equiv p(1)=4\pi T\,.
\label{hat-omega-q}
\ffa
And then combining Eqs. (\ref{Ma-AtI}) and (\ref{Ma-AtII}), we can have a decoupled equation of motion for $A_t(u,\hat{\bf{q}})$ as
\fa
A'''_t+\Big(\frac{f'}{f}-\frac{X'_1}{X_1}+2\frac{X'_3}{X_3}\Big)A''_t
+\Big(-\frac{\mathfrak{p}^2\hat{q}^2X_1}{fX_3}+\frac{\mathfrak{p}^2\hat{\omega}^2X_1}{f^2X_5}+\frac{f'X'_3}{fX_3}-\frac{X_1'X_3'}{X_1X_3}+\frac{X_3''}{X_3}\Big)A'_t=0\,.
\label{Ma-At-dec}
\ffa
For the dual EM theory, the equations of motion can be obtained by setting $A_{\mu}\rightarrow B_{\mu}$ and $X_i\rightarrow \widehat{X}_i$ in the above equations.

Next we discuss the bound of $\gamma$ imposed by the causality and the instabilities.
Note that since Eq. (\ref{Ma-AtI}) gives the relation between $A'_x$ and $A'_t$,
there are only two independent vector modes $A_t$ and $A_y$ and we only need to consider the corresponding equations (\ref{Ma-At-dec}) and (\ref{Ma-Ay}).
It is convenient to formulate Eqs. (\ref{Ma-At-dec}) and (\ref{Ma-Ay}) into the Schr\"odinger form.
To this end, we make the change of variables $dz/du=\mathfrak{p}/f$ and write $A_i(u)=G_i(u)\psi_i(u)$ where we denote $A_{\bar{t}}(u):=A'_t(u)$ and $i=\bar{t},y$.
And then we have
\fa
-\partial_z^2\psi_i(z)+V_i(u)\psi_i(z)=\hat{\omega}^2\psi_i(z)\,,
\label{Sch-form}
\ffa
where $V_i(u)$ is the effective potential. We decompose it into the momentum dependent part and the independent one
\fa
V_i(u)=\hat{q}^2V_{0i}(u)+V_{1i}(u)\,,
\label{Vi-V0-V1}
\ffa
where \cite{Witczak-Krempa:2013aea}
\fa
&&
V_{0\bar{t}}=f\frac{X_1}{X_3}\,,
\,\,\,\,\,\,\,\,\,\,
V_{0y}=f\frac{X_3}{X_1}\,,
\
\\
&&
V_{1\bar{t}}=\frac{f}{4\mathfrak{p}^2X_1^2}[3f(X_1')^2-2X_1(fX_1')']\,,
\
\\
&&
V_{1y}=\frac{f}{4\mathfrak{p}^2X_1^2}[-f(X_1')^2+2X_1(fX_1')']\,.
\ffa
There is a simple relation between $V_{\bar{t}}$ and $V_{y}$ as $V_{\bar{t}}=V_y|_{X_i\rightarrow \widehat{X}_i}$ and vice-versa \cite{Witczak-Krempa:2013aea}.
At the same time, from Eq. (\ref{Xin}), one has $\widehat{X}_i\approx X_i|_{\gamma\rightarrow -\gamma}$ for small $\gamma$.
Therefore, we mainly focus on the discussion of $V_{\bar{t}}$ in what follows.

\begin{figure}
\center{
\includegraphics[scale=0.55]{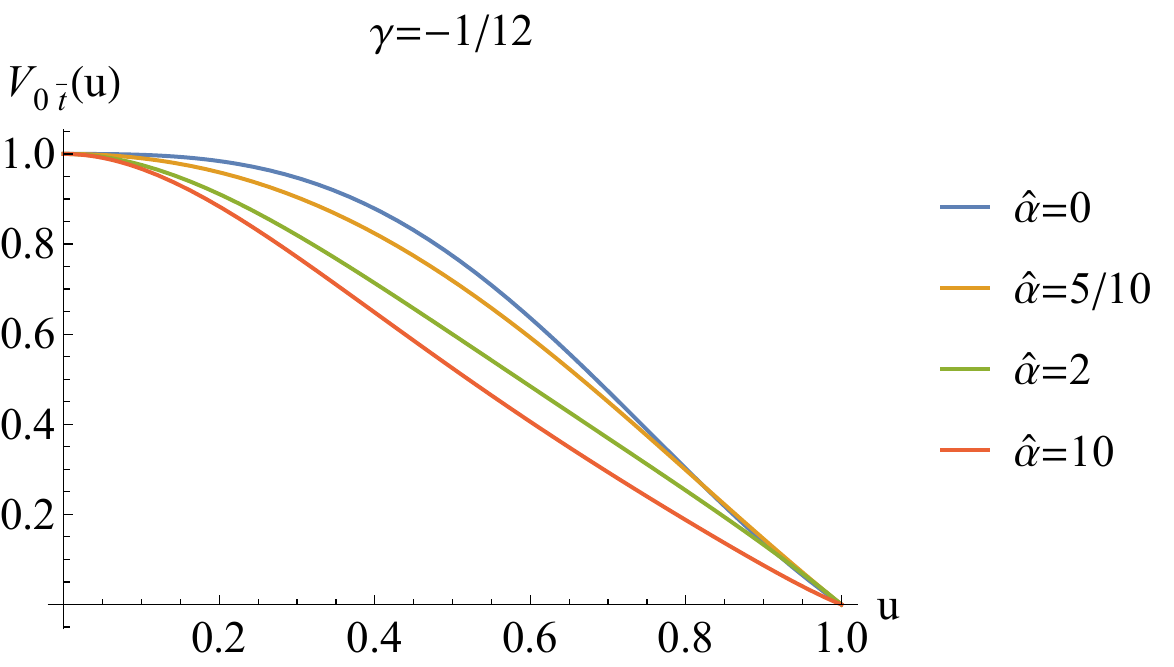}\ \hspace{0.8cm}
\includegraphics[scale=0.55]{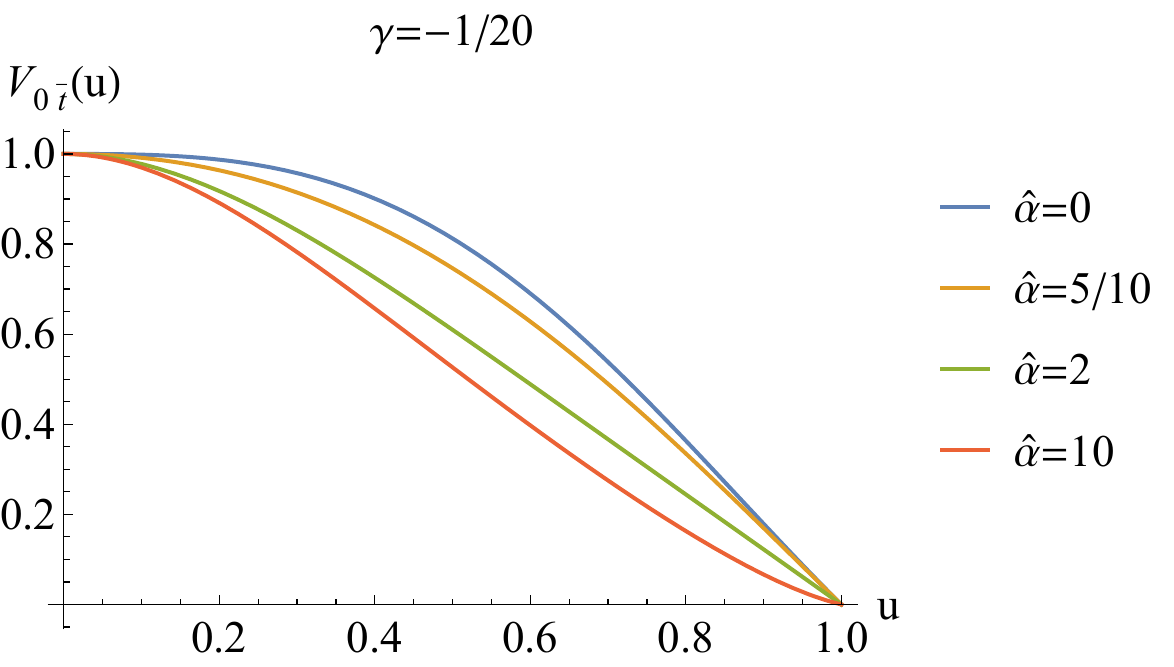}\ \\
\includegraphics[scale=0.55]{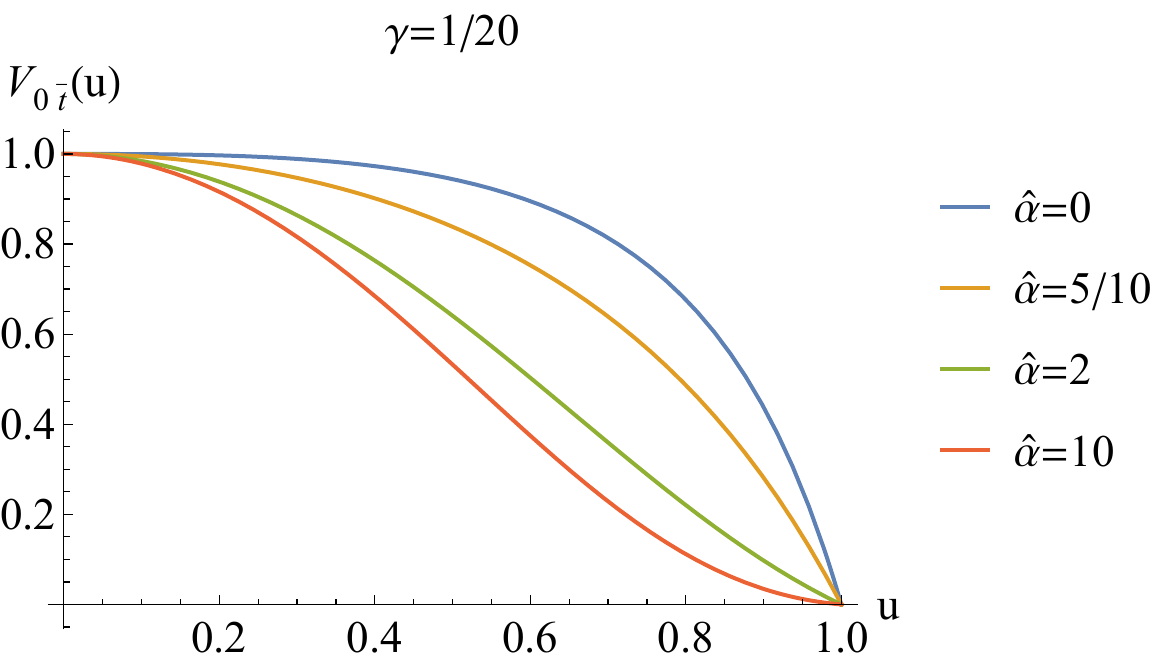}\ \hspace{0.8cm}
\includegraphics[scale=0.55]{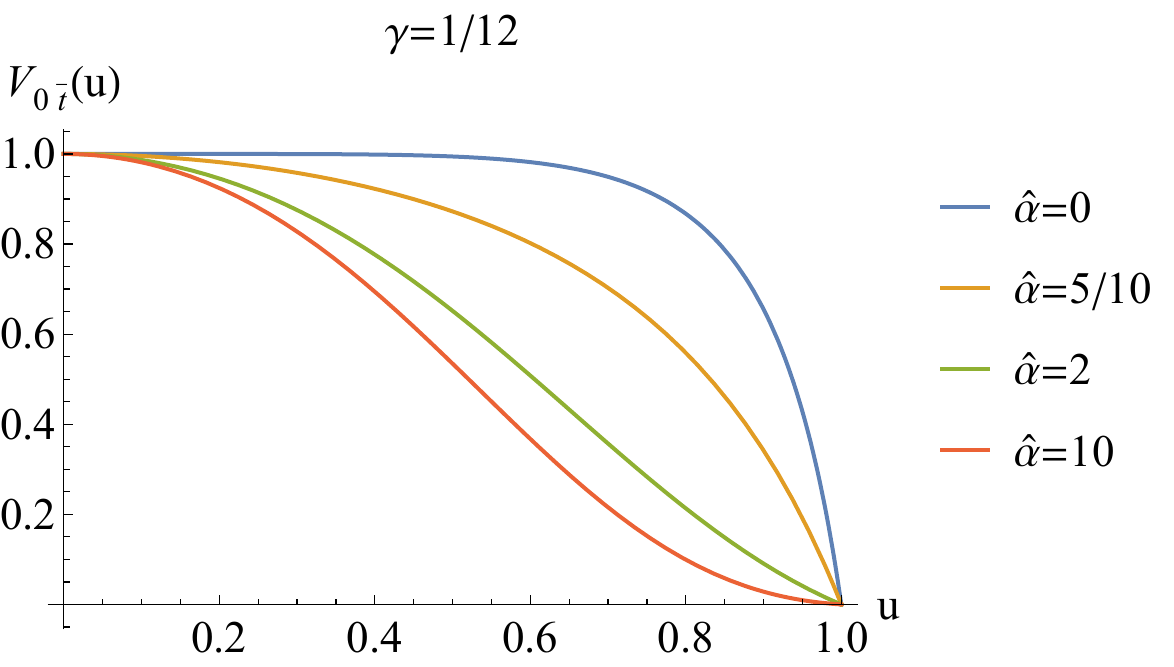}\ \\
\caption{\label{fig-V0-1o12} The shape of the potentials $V_{0\bar{t}}$ of the longitudinal mode $A_t$ for various value of $\gamma\in\mathcal{S}_0$ and $\hat{\alpha}$ is shown.
We find that $V_{0\bar{t}}$ well belongs to the region (\ref{V0i-constraint}).}}
\end{figure}

\begin{figure}
\center{
\includegraphics[scale=0.6]{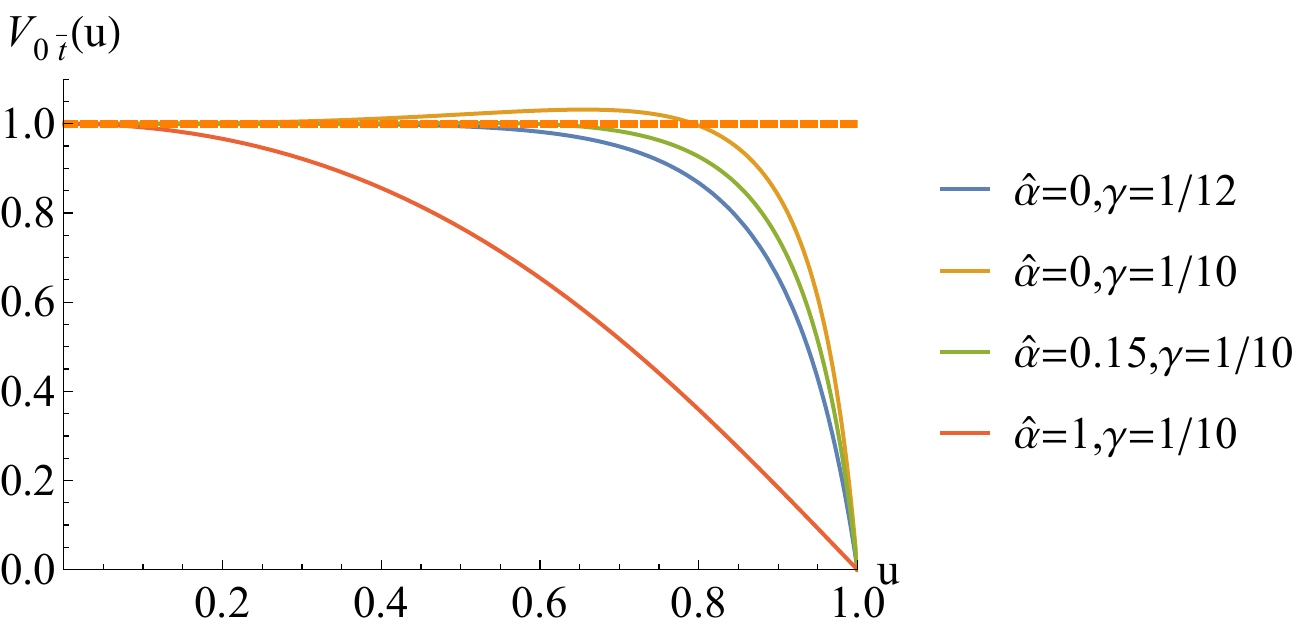}\ \\
\caption{\label{fig-V0t-gamma_1o10} The shape of the potentials $V_{0\bar{t}}$ of the longitudinal mode $A_t$ with momentum dissipation ($\hat{\alpha}\neq0$)
and without momentum dissipation ($\hat{\alpha}=0$) is shown for the $\gamma$ beyond $\mathcal{S}_0$.
For comparision, $V_{0\bar{t}}$ with $\hat{\alpha}=0$ and $\gamma=1/12$ is also plotted.
It clearly show that when the momentum dissipation is introduced, the constraint (\ref{V0i-constraint})
can also satisfied for wider region of $\gamma$ beyond $\mathcal{S}_0$.}}
\end{figure}

Subsequently, we mainly examine whether the constraint $\gamma\in\mathcal{S}_0$ in SS-AdS geometry holds
when the momentum dissipation is introduced. For the $\gamma$ beyond $\mathcal{S}_0$, we present brief comments.
First and foremost we consider the case of the limit of large momentum ($\hat{q}\rightarrow\infty$).
In this limit, the constraint on $V_{0i}$ should be imposed as
\fa
0\leq V_{0i}(u)\leq 1\,.
\label{V0i-constraint}
\ffa
The upper bound of $V_{0i}(u)$ comes from the constraint of the causality in the dual boundary theory \cite{Buchel:2009tt,Brigante:2008gz},
which is a key constraint on the coupling $\gamma$. Otherwise, there will be super-luminal modes with $\omega/q>1$ in this neutral plasma.
While the lower bound of $V_{0i}(u)$ is from the requirement of stability of the vector modes
since in the WKB limit, a negative potential will results in bound states with a negative effective energy,
which corresponds to unstable quasinormal modes in the bulk theory \cite{Myers:2007we}.
FIG.\ref{fig-V0-1o12} shows the shape of the potential $V_{0\bar{t}}$ of the longitudinal mode $A_t$ for various value of $\gamma\in\mathcal{S}_0$ and $\hat{\alpha}$.
It implies that $V_{0\bar{t}}$ well belongs to the region (\ref{V0i-constraint}).
The similar result is found for the potential $V_{0y}$ of the transverse mode $A_{y}$.
Further careful examination indicates that provided $\gamma\in\mathcal{S}_0$
the constraint (\ref{V0i-constraint}) is well satisfied for arbitrary $\hat{\alpha}$.
In fact, when the momentum dissipation is introduced, the constraint (\ref{V0i-constraint})
can be satisfied for wider region of $\gamma$ beyond $\mathcal{S}_0$ (see FIG.\ref{fig-V0t-gamma_1o10}).


\begin{figure}
\center{
\includegraphics[scale=0.6]{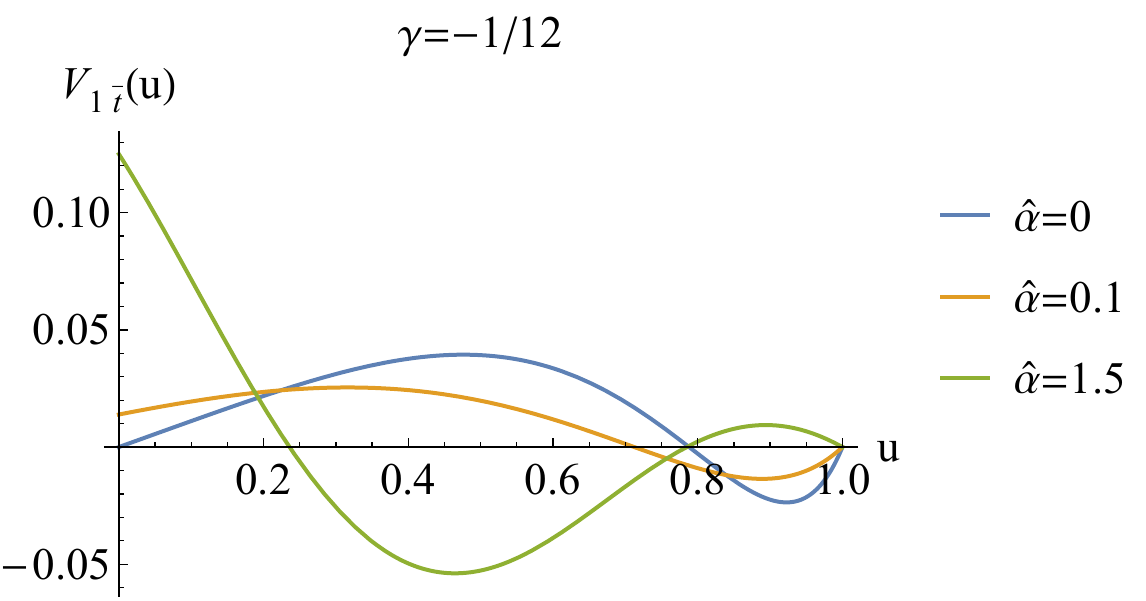}\ \hspace{0.8cm}
\includegraphics[scale=0.6]{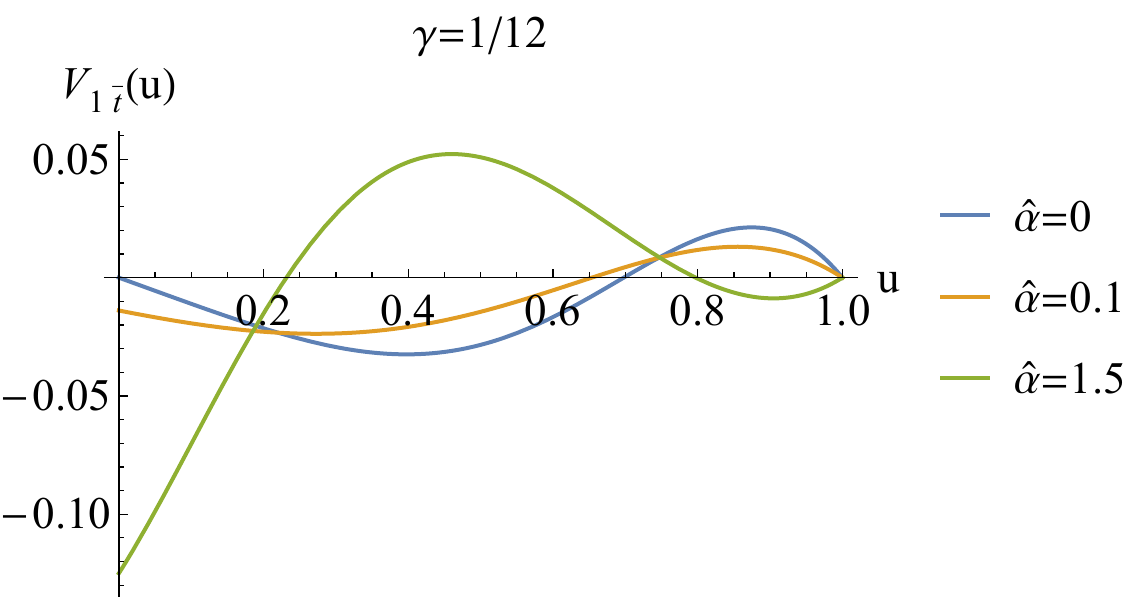}\ \\
\caption{\label{fig-V1-1o12} The shape of the potentials $V_{1\bar{t}}$ with $\gamma=-1/12$ (left plots)
and $\gamma=1/12$ (right plots) for various value of $\hat{\alpha}$ is shown.}}
\end{figure}

\begin{figure}
\center{
\includegraphics[scale=0.55]{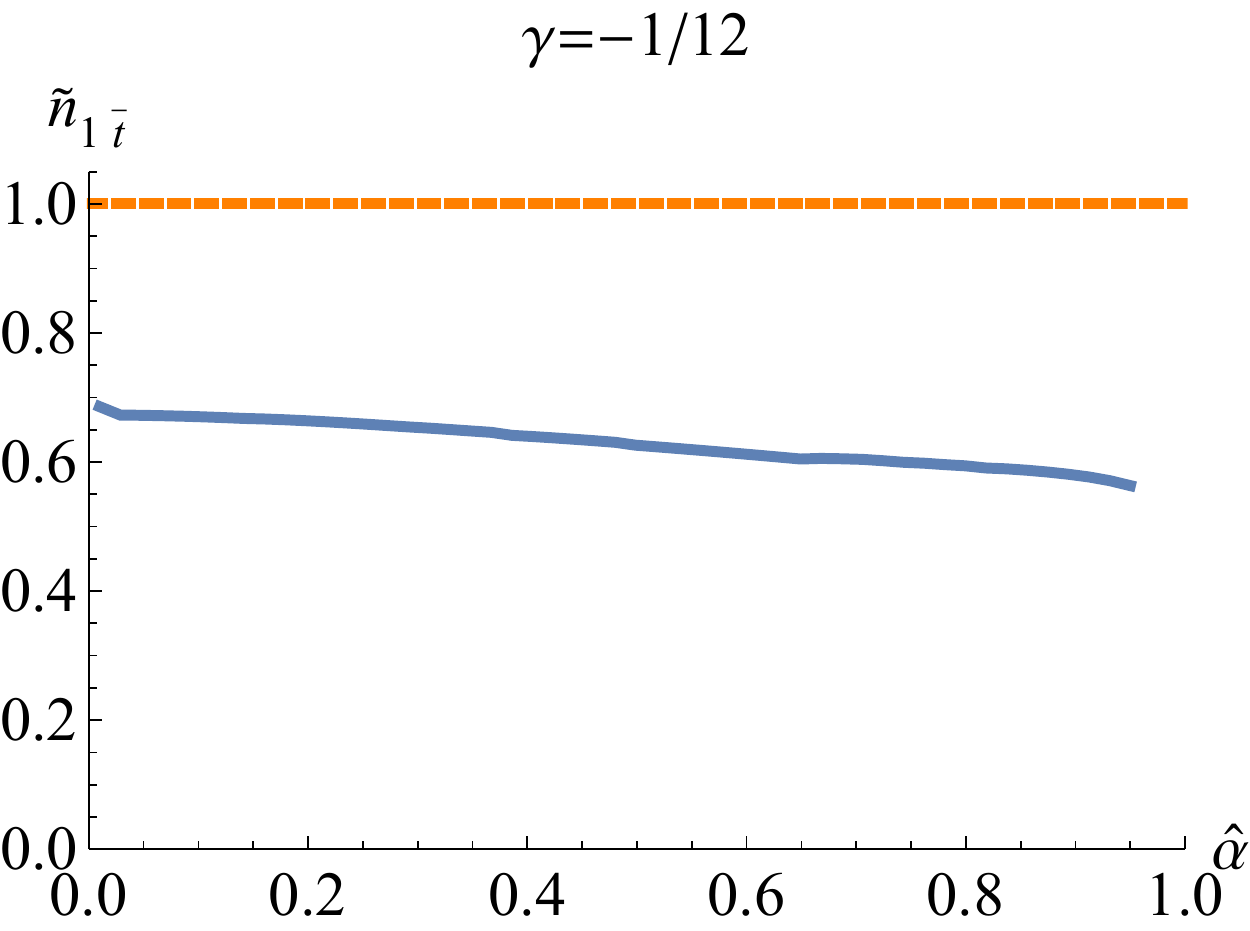}\ \hspace{0.8cm}
\includegraphics[scale=0.55]{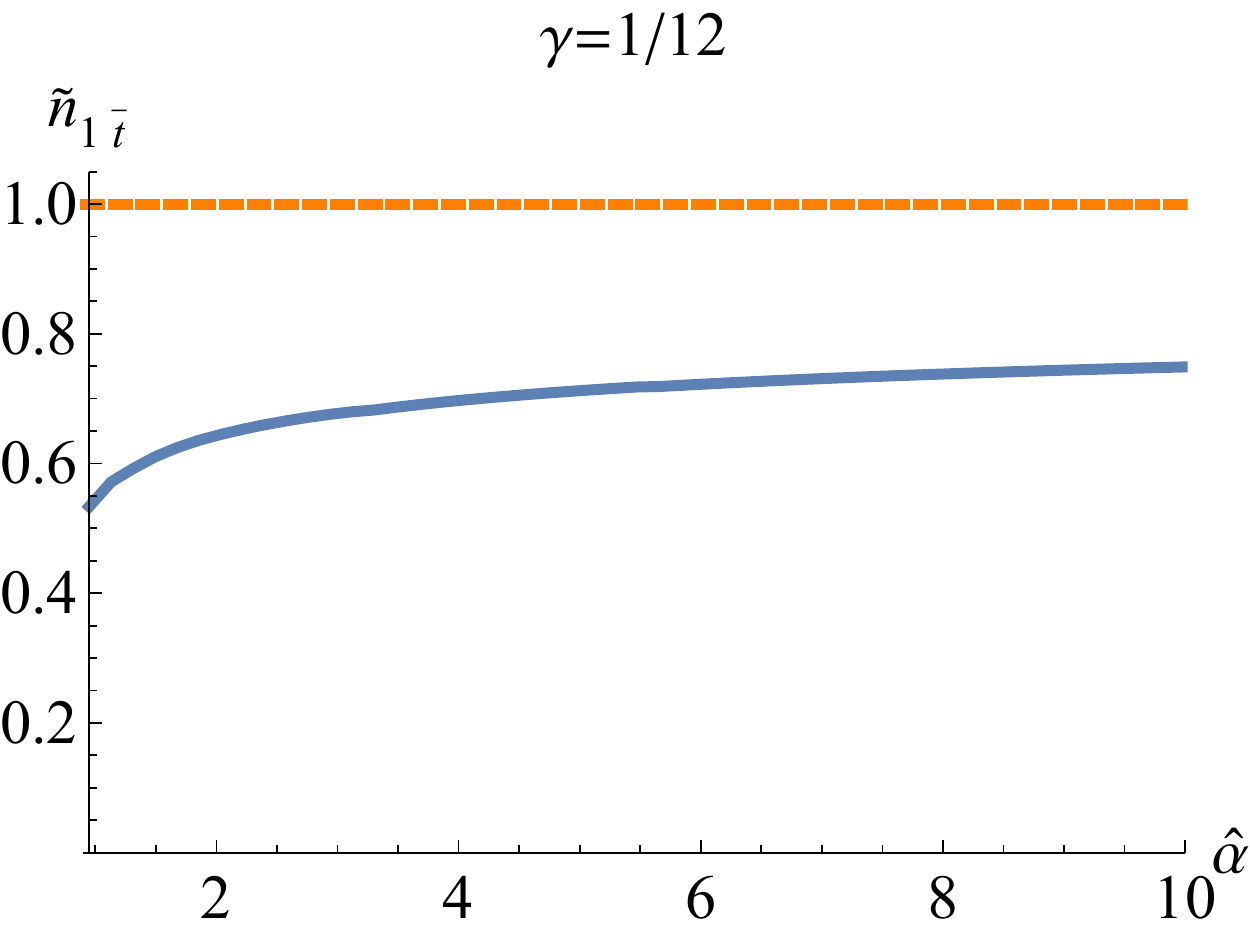}\ \\
\caption{\label{fig-n1tvsa}$\tilde{n}_{1\bar{t}}$ as the function $\hat{\alpha}$ for given $\gamma$ in the region
of $\hat{\alpha}$ in which a negative potential develops close to horizon.
These plots clearly exhibit that $\tilde{n}_{1\bar{t}}$ is always less than unit.}}
\end{figure}

Second we consider the case in the small momentum region, in which $V_{1i}$ play an important role in the effective potential $V_i$.
FIG.\ref{fig-V1-1o12} shows the shape of the potentials $V_{1\bar{t}}$ with $\gamma=-1/12$
and $\gamma=1/12$ for various value of $\hat{\alpha}$.
By careful examination, we find that for $\gamma=-1/12$, $V_{1\bar{t}}(u)$ develops a negative minimum close to the horizon in the region $\hat{\alpha}\in(0,0.95)$,
which means some unstable modes. While for $\gamma=1/12$, the negative minimum in $V_{1\bar{t}}(u)$ appears in $\hat{\alpha}\in(0.95,+\infty)$.
However, although in small momentum region, the potential $V_{1i}$ develops a negative minimum close to the horizon for some regions of $\gamma$ and $\hat{\alpha}$,
we find that there are no unstable modes in these regions by analyzing the zero energy bound state in the potential $V_{1i}$.
We shall demonstrate it below.
As analyzed in \cite{Myers:2007we}, there is a zero energy bound state in the potential $V_{1i}$ by the WKB approximation,
\fa
\label{zero-erergy-bound}
\Big(n-\frac{1}{2}\Big)\pi
=\int_{u_0}^{u_1}\frac{\mathfrak{p}}{f(u)}\sqrt{-V_{1i}(u)}du
\,,
\ffa
where $n$ is a positive integer.
The integration is over the values of $u$ for which the potential well is negative.
Defining $I_i\equiv\big(n-1/2\big)\pi$ and introducing $\tilde{n}_{i\bar{t}}=I_i/\pi+1/2$,
we plot $\tilde{n}_{1\bar{t}}$ as the function $\hat{\alpha}$ for given $\gamma$ in the region
of $\hat{\alpha}$ in which a negative potential develops close to horizon (see FIG.\ref{fig-n1tvsa}).
From this figure, we find that $\tilde{n}_{1\bar{t}}$ is always less than unit
and so no unstable modes appear in the small momentum region.


\begin{figure}
\center{
\includegraphics[scale=0.6]{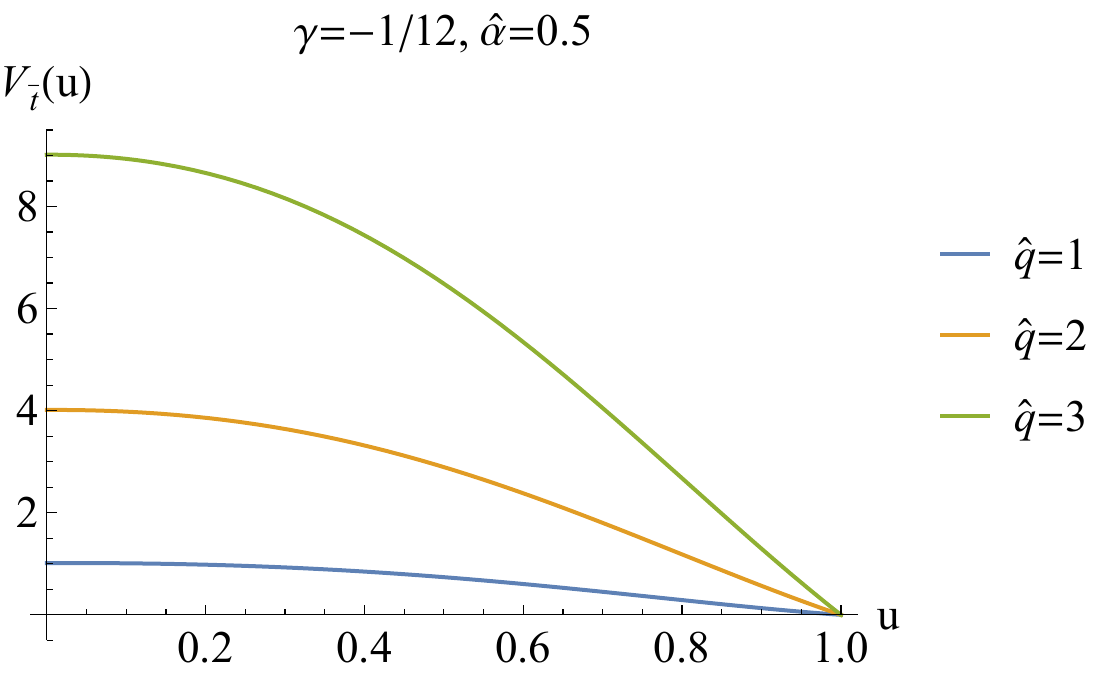}\ \hspace{0.8cm}
\includegraphics[scale=0.6]{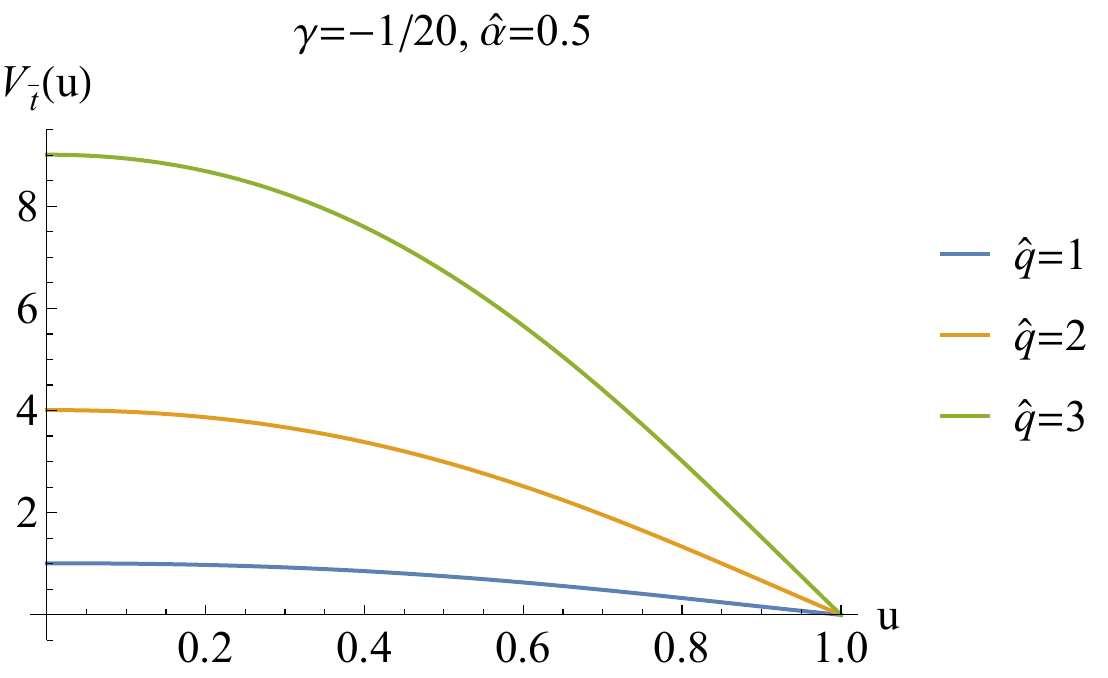}\ \\
\includegraphics[scale=0.6]{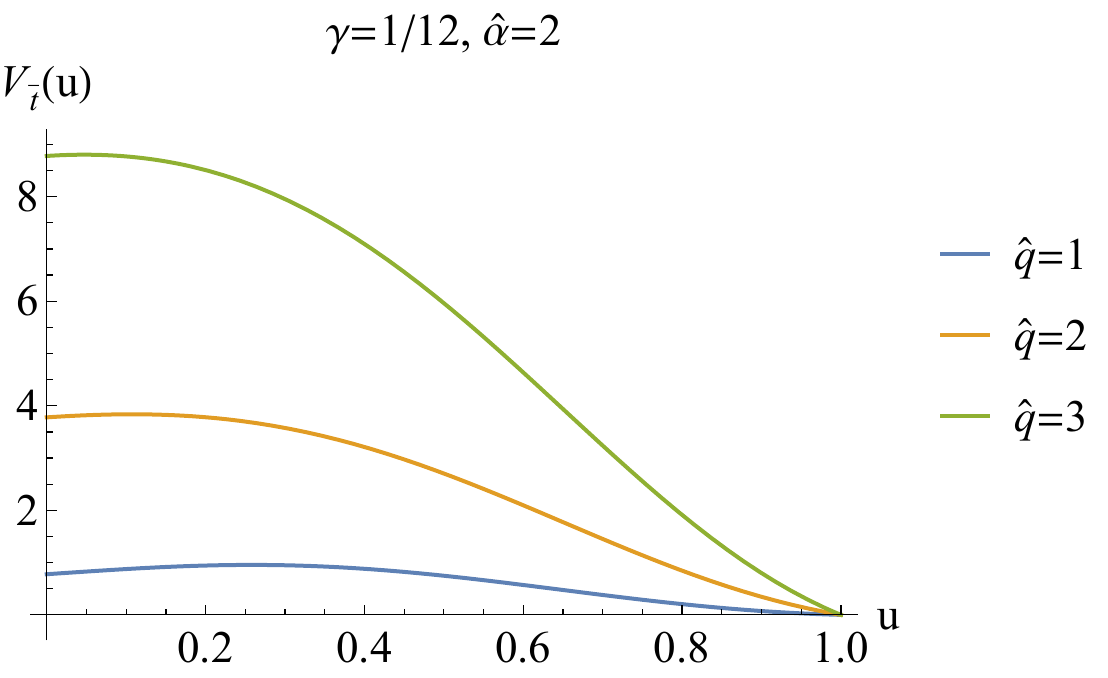}\ \hspace{0.8cm}
\includegraphics[scale=0.6]{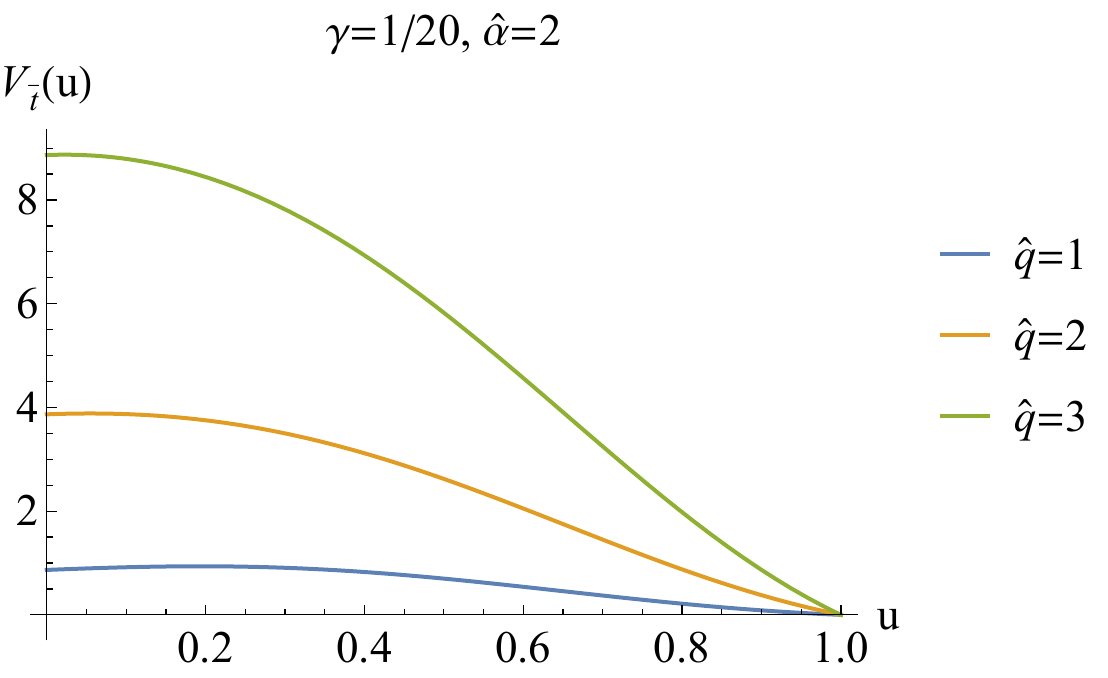}\ \\
\caption{\label{fig-Vt} The potentials $V_{\bar{t}}(u)$ with different $\gamma$ and $\hat{\alpha}$ at some finite momentum.}}
\end{figure}

Finally, after examining the instabilities for small and large momentum limit, we examine the instabilities for some finite momentum.
FIG.\ref{fig-Vt} shows the potentials $V_{\bar{t}}(u)$ with different $\gamma$ and $\hat{\alpha}$ at some finite momentum.
We see that the potential is always positive, which indicates that no unstable modes appear even for the finite momentum.
It is because the positive contribution of $V_{0\bar{t}}(u)$ is larger than the negative one of $V_{1\bar{t}}(u)$.

We conclude that the region $\gamma\in\mathcal{S}_0$ is still physically viable even introducing the momentum dissipation.
In fact, this physically viable region maybe become larger in this neutral axionic geometry (\ref{bl-br}) (see for example FIG.\ref{fig-V0t-gamma_1o10}).
More detailed exploration will be left for the future and here we only restrict ourself in the region $\gamma\in\mathcal{S}_0$.

\end{appendix}

\end{document}